\documentclass[aps,prb,twocolumn,amsmath,amssymb,nofootinbib,superscriptaddress,floatfix]{revtex4}
\usepackage{graphicx,color}
\usepackage{color} 

\usepackage{bm} % bold math
\usepackage{graphicx}% Include figure files
\usepackage{amssymb}
\usepackage{amsmath}
\usepackage{dcolumn} % Align table columns on decimal point

\usepackage[hypertex]{hyperref}

\begin{document}

%draft${}_{-}$3dtop${}_{-}$chiral${}_{-}$bdg${}_{-}$08${}_{-}$04${}_{-}$08c${}_{-}$.tex

\title{
Classification of topological insulators and 
superconductors in three spatial dimensions
}

\author{Andreas P.\ Schnyder}
\affiliation{Kavli Institute for Theoretical Physics,
  University of California,
  Santa Barbara,
  CA 93106,
  USA}

\author{Shinsei Ryu}
\affiliation{Kavli Institute for Theoretical Physics,
  University of California,
  Santa Barbara,
  CA 93106,
  USA}

\author{Akira Furusaki}
\affiliation{Condensed Matter Theory Laboratory,
             RIKEN,
             Wako,
             Saitama 351-0198,
             Japan}

\author{Andreas W.\ W.\ Ludwig}
\affiliation{Department of
             Physics, University of California,
             Santa Barbara, CA 93106, USA}

\date{\today}

\begin{abstract}

We systematically study topological phases of
insulators and
superconductors (or superfluids)
in three spatial dimensions (3D).
We find that there exist 3D topologically non-trivial
insulators or superconductors in 
five out of ten symmetry classes introduced
in seminal work by Altland and Zirnbauer within the context of
random matrix theory, more than a decade ago. 
One of these is the recently introduced $\mathbb{Z}_2$ topological insulator
in the symplectic (or spin-orbit) symmetry class.
We show there exist precisely four more topological insulators.
For these systems, 
all of which are time-reversal invariant in 3D,
the space of insulating ground states satisfying
certain discrete symmetry properties
is partitioned into topological sectors that are
separated by quantum phase transitions.
Three of the above five
topologically
non-trivial phases can be realized 
as time-reversal invariant superconductors,
and in these the different topological sectors
are characterized by an integer winding number defined in momentum space.
When such 3D topological insulators are terminated by a two-dimensional 
surface,
they support a number
(which may be an arbitrary non-vanishing even number for singlet pairing)
of Dirac fermion 
(Majorana fermion when spin rotation symmetry is completely broken)
surface modes which remain gapless under arbitrary perturbations
of the Hamiltonian 
that  preserve
the characteristic 
discrete symmetries, including disorder. 
In particular, these surface modes completely
evade Anderson localization from random impurities.
These topological phases can be thought of as three-dimensional 
analogues of well known paired topological phases in two
spatial dimensions such as
the spinless chiral $(p_x\pm {i}p_y)$-wave superconductor
(or Moore-Read Pfaffian state).
In the corresponding topologically non-trivial 
(analogous to ``weak pairing'') and topologically trivial 
(analogous to ``strong pairing'') 3D phases,
the wave functions exhibit markedly distinct behavior.
When an electromagnetic U(1) gauge field
and fluctuations of the gap functions are included in the dynamics,
the superconducting phases with 
non-vanishing winding number possess
non-trivial topological ground state degeneracies.
\end{abstract}

\maketitle

%%%%%%%%%%%%%%%%%%%%%%%%%%%%%%%%%%%%%%%%%%%%%%%%%%%%%%%%%%%%%%%%%%%
\section{Introduction}
\label{sec: introduction}

\begin{table*}[t]
\begin{center}
\begin{tabular}{|c|c||c|c|c||c|c|c|}\hline
&  & TRS   & PHS  & SLS  & $d=1$ & $d=2$ & $d=3$ 
\\\hline\hline
standard       &A (unitary)   &$0$  &$0$  &$0$ &   -            &$\mathbb{Z}$   & - \\ \cline{2-8}
(Wigner-Dyson) &AI  (orthogonal) &$+1$ &$0$  &$0$ &     -          &     -          &- \\ \cline{2-8}
               &AII (symplectic) &$-1$ &$0$  &$0$ &    -           &$\mathbb{Z}_2$ & $\mathbb{Z}_2$\\ \hline\hline
chiral         &AIII (chiral unitary)&$0$  &$0$  &$1$ &$\mathbb{Z}$   &     -          & $\mathbb{Z}$\\ \cline{2-8}
(sublattice)   &BDI (chiral orthogonal) &$+1$ &$+1$ &$1$ &$\mathbb{Z}$   &      -         & - \\ \cline{2-8}
               &CII  (chiral symplectic) &$-1$ &$-1$ &$1$ &$\mathbb{Z}$   &       -        &$\mathbb{Z}_2$\\\hline \hline
BdG            &D    &$0$  &$+1$ &$0$ &$\mathbb{Z}_2$ &$\mathbb{Z}$   & - \\ \cline{2-8}
               &C    &$0$  &$-1$ &$0$ &       -        &$\mathbb{Z}$   & -\\ \cline{2-8}
               &DIII &$-1$ &$+1$ &$1$ &$\mathbb{Z}_2$ &$\mathbb{Z}_2$ &$\mathbb{Z}$\\ \cline{2-8}
               &CI   &$+1$ &$-1$ &$1$ &       -        &         -      & $\mathbb{Z}$\\ \hline
\end{tabular}
\caption{
\label{tab: rmt}
Ten symmetry classes of single particle Hamiltonians classified in terms of 
the presence or absence of time-reversal symmetry (TRS) and
particle-hole symmetry (PHS), 
as well as  sublattice (or ``chiral'') symmetry (SLS).
\cite{Zirnbauer96, Altland97}
In the table, the absence of symmetries is denoted  by ``$0$''. 
The presence of these symmetries is denoted either by ``$+1$'' or
``$-1$'', depending on whether the (antiunitary) operator implementing the 
symmetry at the level of the single-particle Hamiltonian 
squares to ``$+1$''or ``$-1$'' (see text). 
[The index $\pm 1$ equals $\eta_c$ in Eq.\ (\ref{eq: def epsilon and eta});
here $\epsilon_c=+1, -1$ for TRS and PHS, respectively.]
For the first six entries of the TABLE (which can be realized in non-superconducting
systems) TRS $=+1$ when the SU(2) spin is integer
[called TRS (even) in the text]
and
TRS $=-1$ when it is a half-integer 
[called TRS (odd) in the text].
For the last four entries, the superconductor 
``Bogoliubov-de Gennes'' (BdG) symmetry classes D, C, DIII, and CI,
the Hamiltonian preserves SU(2) spin-1/2 rotation symmetry when
PHS=$-1$ 
[called PHS (singlet) in the text],
while it does not preserve SU(2) 
when PHS=$+1$ [called PHS (triplet) in the text].
The last three columns list all topologically non-trivial
quantum ground states as a function of symmetry
class and spatial dimension.
The symbols $\mathbb{Z}$ and $\mathbb{Z}_2$ indicate
whether the space of quantum ground states
is partitioned into topological sectors labeled by an integer or
a $\mathbb{Z}_2$ quantity, respectively.
}
\end{center}
\end{table*}

Quantum states of matter are characterized
not only by the structure of the energy spectrum
but also by the nature of wave functions.
\cite{Wen90,Wen92,Wenbook}
Of particular importance are \textit{topological} properties of 
wave functions, i.e., 
properties that are invariant under small adiabatic deformations of the
Hamiltonian.
A classic example of such topological characteristics is the quantized 
Hall conductivity $\sigma_{xy}$ in the integer quantum Hall effect (IQHE), 
which occurs at low temperature and high magnetic field
in two-dimensional (2D) electronic systems with broken
time-reversal symmetry (TRS). The transverse (Hall) conductivity $\sigma_{xy}$, 
arising from topologically protected edge currents,
can be interpreted as an integer Chern number
(TKNN integer),\cite{Klitzing, Thouless82,Kohmoto} 
a quantized topological invariant which characterizes the different
topological ground states.

In the past few years, it has been realized that topological phases
supporting topologically protected 
states appearing at the sample boundaries
can also exist 
in two- and three-dimensional \emph{time-reversal invariant}
systems in the absence of an external magnetic 
field.\cite{KaneMele, Roy06, Moore06, Roy3d,Fu06_3Da,Fu06_3Db,Fu06,Bernevig05,Murakami06,Xiao-LiangQi08}
These topological states occur in certain materials with
a bulk band gap generated by strong spin-orbit interactions and are known
as $\mathbb{Z}_2$ topological insulators.
Unlike the integer quantum Hall states, the two-dimensional version
of the $\mathbb{Z}_2$ topological insulator,
which has been dubbed ``quantum spin Hall'' (QSH) state,
does not carry any net charge current along the edges.
Instead, when a U(1) part of the SU(2) spin rotation symmetry is
conserved, electrons with opposite spin propagate in opposite directions, 
which gives rise to a quantized spin Hall conductance.
\cite{KaneMele,Bernevig05}
Remarkably, the topological order of the QSH state survives even under
a (small) breaking of the full spin-rotation symmetry.
Consequently, the QSH insulator cannot be characterized by a quantized spin
Hall conductivity.
Rather, as shown by Kane and Mele in Ref.\ \onlinecite{KaneMele},
there is a $\mathbb{Z}_2$ topological invariant that classifies the
topological properties of the QSH states in a similar way as the Chern number
does in the IQHE.
A simple interpretation of the $\mathbb{Z}_2$ invariant
is in terms of doublets of edge modes: in the QSH 
phase the edge states consist of an odd number of Kramers
doublets, whereas the conventional
band insulator is characterized by 
an even number
(including zero)
of pairs of edge states.
The odd number of Kramers doublets is robust against disorder 
\cite{Takane04,Zirnbauer92,Ando98,CongjunWu06,Xu06,obuse2007,essin2007}
and interactions.
\cite{CongjunWu06,Xu06}
In particular, when subject to time-reversal invariant random impurity potentials,
there is always one perfectly conducting channel,
as long as the bulk topological properties are not altered by disorder.

There is a natural generalization of the $\mathbb{Z}_2$
topological insulator to three dimensions (3D).
\cite{Fu06_3Da,Fu06_3Db,Moore06,Roy3d} 
Similar to the 2D version, there are now
four independent $\mathbb{Z}_2$ topological invariants
which describe the number of Kramers degenerate band crossings 
(Dirac points) in the spectrum on the surface of the 3D bulk,
thereby distinguishing the conventional insulator, the
topologically trivial phase from the topologically non-trivial phase.
Although the effects of disorder and interactions on the
$\mathbb{Z}_2$ topological insulator have
been less well studied in 3D than in the 2D case,
there are known to exist
gapless surface modes 
in the topologically non-trivial 3D phase 
which
are robust against arbitrary strong disorder 
as long as the latter does not alter the bulk topological properties,
in analogy to the QSH effect (QSHE) in 2D. 
\cite{Fu06_3Db, Ando98, Ostrovsky07, Ryu07, Bardarson07,Nomura07}
These delocalized surface states, whose Fermi surface
encloses an odd number of Dirac points,
form a two-dimensional ``$\mathbb{Z}_2$ topological metal''.
\cite{Fu06_3Db,Nomura07,Haldane04}

Recently, a series of experiments
have been performed on 
certain
candidate materials for  
$\mathbb{Z}_2$ topological insulators.
For example, the QSH effect has been observed in
HgTe/(Hg,Cd)Te semiconductor quantum wells.
\cite{Xiao-LiangQi05,Bernevig-Taylor-Zhang06, Dai07,Koenig07,Koenig08}
Moreover, a 3D $\mathbb{Z}_2$ topological phase
has been predicted for strained
HgTe and 
for
Bismuth-Antimony alloys.
\cite{Fu06_3Db, Dai07,Fukui06}
Indeed, photoemission experiments on the latter system have
revealed an odd number of Dirac points inside 
the Fermi surface on the (111)-surface, thereby providing (indirect) evidence
for the existence of a non-trivial topological phase
in three spatial dimensions.
\cite{Fu06_3Db,Hasan07}

In this paper we provide an {\it exhaustive} classification of
topological insulators and superconductors.
Our classification is for non-interacting systems of fermions.
However, since there is a gap, our results also apply to interacting
systems as long as the strength of the interactions is sufficiently
small as compared to the gap.
As the majority of previous works studied two-dimensional topological phases, 
we shall be mostly concerned with the classification of 3D systems,
and only briefly comment 
on one- and two-dimensional topological insulators in the discussion section
(Sec.\  \ref{sec: discussions}).
In the same spirit as in the treatments of
$\mathbb{Z}_2$ topological insulators,
we impose several discrete symmetries
on a family of quantum ground states.
We then ask if different quantum states can be transmuted
into each other, without crossing a quantum phase transition,
by a continuous deformation respecting the discrete symmetries.

If we are to include spatially inhomogeneous 
deformations of quantum states,
such as those arising, e.g., from
the presence of random impurity potentials, 
the natural discrete symmetries we should think of
would be those considered in the context of disordered systems.
\cite{footnote1}
It is at this stage that we realize that
the existence of the classification of \textit{random} Hamiltonians,
familiar from the theory of {\it random matrices},
will become very useful for this purpose.

Specifically, following Zirnbauer, and Altland and Zirnbauer (AZ),
\cite{Zirnbauer96, Altland97} 
all possible symmetry classes of random matrices,
which can be interpreted as a Hamiltonian of some
non-interacting fermionic system,
can be systematically enumerated:
there are ten symmetry classes in total.
(For a summary, see Table \ref{tab: rmt}.)
The basic idea  as to why  there are precisely ten is easy
to understand. Roughly, the only  generic symmetries
relevant for any system are time-reversal symmetry (TRS),
and charge conjugation or particle-hole symmetry (PHS).
%\textcolor{blue}{
%Both are represented on the Hilbert space
%by anti-unitary operators, and can be written in the form ${\hat U} {\hat K}$, 
%with ${\hat K}=$ complex conjugation, and $ {\hat U}=$
%unitary.
%}
Both can be represented by antiunitary operators
on the Hilbert space on which the single-particle Hamiltonian
(a matrix) acts, 
and can be written 
\cite{footnote_PH}
on this space 
in the form $K U$, 
with $K=$ complex conjugation, and $U=$
unitary.
Any of these two symmetries can either be absent,
which we denote by $0$, or be present and square to
the identity operator, or to minus the identity operator,
which we denote by  $+1$ or $-1$, respectively.
This gives nine possible 
choices for the pair of
symmetries TRS and PHS (Table \ref{tab: rmt}).
However, since we consider TRS and PHS, we may also consider,
in addition, their product SLS $:= $ TRS $\times$ PHS, often
referred to as ``sublattice'' (or ``chiral'') symmetry.
Now, for eight of the nine assignments of 
a pair of $0,\pm 1$ to
the pair of symmetries TRS and PHS, the presence or
absence of the product  SLS of these symmetries is uniquely determined (Table \ref{tab: rmt}).
But the assignment
$({\rm TRS,PHS} )= (0,0)$ allows for SLS to 
either be present (SLS $=1$)
or absent (SLS $=0$). Therefore one obtains
ten symmetry classes (Table \ref{tab: rmt}),
an exhaustive list.

The so-obtained
ten (AZ) symmetry classes of random matrices are
conventionally
named after the mathematical classification of symmetric spaces,
and called A, AI, AII, AIII, BDI, CII, D, C, DIII, and CI
(Table \ref{tab: rmt}).
This AZ classification includes
the three previously known,\cite{MehtaRandomMatrices}
so-called ``Wigner-Dyson symmetry classes''
(or ``standard symmetry classes'')
relevant for the physics of Anderson localization
of electrons in disordered solids
[corresponding to orthogonal (AI), unitary (A),
and symplectic (AII) random matrix ensembles].
Three so-called ``chiral classes'' can be obtained,
from the Wigner-Dyson classes,
by imposing an additional SLS; 
these are conventionally called
``chiral orthogonal'' (BDI), ``chiral unitary'' (AIII), and
``chiral symplectic'' (CII) symmetry classes.
A well-known prototypical example of a system in a chiral symmetry class
is a disordered tight-binding model on
a bipartite lattice, such as the random hopping model, and
the random flux model.\cite{Furusaki99}
Finally, there are four additional symmetry classes (D, C, DIII, and CI)
describing the (Anderson-like) localization physics of
the non-interacting 
Bogoliubov-de Gennes (BdG)
quasiparticles existing deep inside the superconducting
state 
of disordered superconductors, as described
within a mean-field treatment of pairing.
(Four symmetry classes arise since SU$(2)$ spin-rotational invariance, 
or TRS may be present or absent.)

In terms of this terminology, the
above mentioned
$\mathbb{Z}_2$ topological insulator is
a topologically non-trivial insulator within the symplectic
(or: ``spin-orbit'') 
symmetry class (class AII).
In this paper,
we pursue this direction further
and provide a classification of {\it all possible}
topological insulators in 3D.
Specifically, we
take the classification by AZ and ask if
different Hamiltonians can be continuously deformed 
into each other within a given symmetry class.

\subsection{Summary of results}

One of the key results of the present paper is our
finding that for five out of the above-mentioned ten (AZ) 
symmetry classes for random matrices, there 
exist topologically non-trivial insulators
\cite{footnoteInsulator}
in three spatial dimensions.
These classes are:
\begin{itemize}
\item
symplectic symmetry class (class AII),
\item
chiral unitary symmetry class (class AIII),
\item
chiral symplectic symmetry class (class CII),
\item
BdG symmetry class of superconductors
with  time-reversal (TR) but no SU(2) spin rotation
symmetry (class DIII),

and
\item
BdG symmetry class
of superconductors
with both TR and SU(2) spin rotation
symmetries (class CI).
\end{itemize}
All these symmetry classes possess a TRS of some 
form.\cite{TRS in class AIII,Foster07,FendleyKonik00} 
Our  result
is summarized in the last column of Table \ref{tab: rmt},
where the symbols
$\mathbb{Z}$ and $\mathbb{Z}_2$ indicate, whether 
the space of quantum ground states 
is partitioned into topological sectors labeled by an integer or
a $\mathbb{Z}_2$ quantity, respectively.

We are going to derive these findings by using
two complementary strategies.
First, we introduce a suitable topological invariant which 
takes on integer values, and can be used
to label topologically distinct quantum ground states (see Sec.~\ref{sec: characterization in the bulk}).
Second, we study two-dimensional boundaries terminating
3D topological insulators and use the appearance of 
gapless surface modes as a 
diagnostic
for the topological
nature of
the 3D bulk
 properties (see Sec.~\ref{sec: characterization at the boundary}). The latter is accomplished
by considering Dirac Hamiltonian representatives of 3D topological insulators.
Before we turn to a more detailed and technical discussion,
we outline below these strategies in general terms.

\subsubsection{Bulk topological invariant}

To characterize topological properties
of bulk wave functions in classes AIII, DIII, and CI,
which are classes that can be realized 
as time reversal (TR) invariant superconductors,
we introduce an integer-valued topological invariant (winding number),
to be denoted by $\nu$.
This winding number can be defined for those symmetry classes
in which the Hamiltonian can be brought into  block off-diagonal form,
a well known property of random matrices in all 
the so-called chiral symmetry classes AIII, BDI, and CII.
This turns out to be also a property of
symmetry classes DIII and CI
(and arises from the presence of a
(``sublattice'', or ``chiral'')
 symmetry which is
a combination of PHS and TRS).
One of the simplest examples of such a quantum
state with a non-trivial winding number is in fact
the well known Balian-Werthamer (BW) state
\cite{balian1963,VolovikBooks}
of the B phase of liquid ${ }^{3}\mathrm{He}$.
In the above language, 
the BdG fermionic quasiparticles
in  ${ }^{3}\mathrm{He}$ B
are in a 3D topological insulating phase in class DIII,
with winding number $\nu=1$.

However, when additional discrete symmetries are present in a given symmetry class,
these can (and will) restrict the possible values of the winding number $\nu$ to a 
subset of the integers.
Indeed, while an arbitrary integral winding number
can be realized in classes DIII and AIII,
only an even winding number turns out to be allowed for class CI.
For classes BDI and CII, on the other hand,
we do not find any example with a non-trivial winding number.

\subsubsection{Surface Dirac/Majorana fermion modes}

As anticipated from the examples of
topological insulators in 2D such as
the IQHE and the $\mathbb{Z}_2$ topological insulators,
the non-trivial topological properties of the
quantum state in the 3D bulk manifest themselves through the 
appearance of gapless
modes at a 2D surface terminating the 3D bulk;
these turn out to be gapless Dirac (or Majorana) fermion
modes.  \cite{Halperin82,Callan85,Fradkin86,Boyanovsky87,Kaplan92,
SalomaaVolovik1988,GrinevichandVolovik1988}
The converse is also true: the physics at the boundary faithfully
reflects (indeed ``\textit{holographically}'')
the non-trivial topological features of the bulk quantum state,
in a way that is reminiscent of the situation familiar from the 
quantum Hall states
\cite{WittenOldChernSimonsCFT,MooreRead91,Halperin82,Hatsugai93} 
[and also from recent work on gravitational Chern-Simons theory in
$(2+1)$ dimensions.
\cite{Witten2007TwoPlusOneGravityAndAdS}]
We are thus led to consider the nature and the properties of Dirac
fermions appearing at two-dimensional surfaces terminating
the 3D bulk {\it as a tool} to characterize and to learn about the 
topological properties of the wave functions of the three-dimensional 
bulk of interest.

Indeed, a complete classification scheme of properties
of Dirac fermions in two spatial dimensions has recently appeared
in the work of Bernard and LeClair (BL),\cite{Bernard01}
and we will use their results extensively.
\cite{NoKnowledgeofBernardLeClair}
Interestingly, and of key importance to our goal,
is the fact that the BL-classification scheme
consists of 13 symmetry classes, 
not just the 10 (AZ) classes mentioned above.
This is due to the ``Dirac structure'' of the Hamiltonian:
in addition to the ordinary ten symmetry classes,
each one of the three classes AIII, DIII and CI
(i.e., classes which can be realized in
TR invariant superconductors),
subdivides in fact into two symmetry classes
(not just one, as in the AZ scheme which applies
to random Hamiltonian not of ``Dirac form'').
The gapless nature of 2D Dirac fermions in 
the extra three symmetry classes turns out to be entirely robust
against any perturbation (respecting the symmetries 
of a given symmetry class),
including those breaking translational invariance 
(i.e., disorder potentials).
Remarkably, it is precisely these extra three classes 
which are realized at a surface of  bulk topological insulators
in classes AIII, DIII, and CI in three spatial dimensions.
More specifically, for symmetry classes AIII and DIII
there exist 3D topological insulators possessing {\it any} number of
gapless 2D surface Dirac and Majorana fermion modes, respectively,
which are stable to arbitrary perturbations (respecting the symmetries).
For class CI, on the other hand,
only an \textit{even}
number of such gapless 2D Dirac fermion modes
which are robust to perturbations can be realized at a surface.
This should be contrasted with the even-odd effect
\cite{Fu06_3Da,Fu06_3Db,Moore06,Roy3d}
governing the robustness of the gaplessness of the spectrum 
of the Dirac fermion modes at a surface of a three-dimensional 
$\mathbb{Z}_2$ topological insulator, a feature which is protected
by the $\mathbb{Z}_2$ invariant.

For class CII the situation is similar to the case of $\mathbb{Z}_2$ 
topological insulators in class AII (the symplectic symmetry class).
A three-dimensional insulator in class CII can support 
surface Dirac fermions that are stable against any symmetry-preserving
perturbations.
We will explicitly demonstrate this below for
the case when the number of surface Dirac fermions is two.
On the other hand, when the number of such  flavors is
twice an even integer, the surface Dirac fermions are not protected
from acquiring a mass. This stability of the gapless
surface Dirac fermions has nothing to do with the winding number
mentioned above.

\subsubsection{Examples and many-body wave functions}

The discussion so far has been solely in terms of single-particle physics.
This is not to say, however, that these results are completely irrelevant to
interacting many-body systems: 
in particular, when viewed as mean-field ground states,
the families of states considered above can naturally arise as a consequence
of strong correlations. For example,
the most interesting topological features of
the Moore-Read Pfaffian state
of the fractional quantum Hall (FQH) effect 
for the half-filled Landau level
can be understood in terms of
the  ground state of a certain ($p_x + {i} p_y$)
BCS superconductor,
within a mean field treatment of pairing.
\cite{Read00}
Moreover, once the dynamical fluctuations of the pairing potential
and of the electromagnetic U(1) gauge field are included,
the superconducting ground state is topologically ordered.
\cite{Diamantini1996,Hansson04, Diamantini2006,Oshikawa06}
In a similar fashion, the 3D topological phases realized
in superconducting classes DIII, AIII and CI
are, when viewed as many-body wave functions, 
3D analogues of paired FQH states such as
the Moore-Read Pfaffian,
\cite{MooreRead91}
the Halperin 331,
\cite{Halperin83}
and
the Haldane-Rezayi states.
\cite{HaldaneRezayi88}
In particular,
the BW state of the B-phase of liquid ${ }^{3}\mathrm{He}$
can be thought of as a 3D analogue of the Moore-Read Pfaffian state.
Below, we will discuss in more detail
the properties 
of real-space wave functions
of topologically non-trivial and
of topologically trivial phases
of these 3D topological insulators,
which are analogues of the
familiar ``weak-pairing'' and the ``strong-pairing''
states, respectively,  of the Moore-Read Pfaffian state in two
spatial dimensions.
We will also discuss topological degeneracies arising in such 3D phases.

\subsection{Outline}

This paper is structured as follows.
Since the symmetry classification \cite{Altland97,Bernard01}
of Hamiltonians which emerged in the context of random
matrix theory is crucial for our discussion,
we give a rather pedagogical description of it
in Sec.\ \ref{sec: symmetry classification of non-interacting Hamiltonians}.
The topological winding number $\nu$ is
introduced in Sec.\ \ref{sec: characterization in the bulk}
to characterize the bulk ground state wave functions.
We have a close look at the surface Dirac fermions
in Sec.\ \ref{sec: characterization at the boundary},
following Bernard and LeClair. \cite{Bernard01}
Explicit examples of 3D Dirac insulators with a minimal number
of surface Dirac fermions are constructed
in Sec.\ \ref{sec: 3D Dirac Hamiltonian}.
To clarify the connection between
the topological properties of the
quantum state in the 3D bulk
and
the appearance of gapless Dirac fermions at 2D surfaces,
we consider a topological field theory description
in terms of
(doubled)
Chern-Simons theory
in Sec.\ \ref{sec: topological field theory description}.
Finally, we discuss the 3D topological insulators as many-body systems
in Sec.\ \ref{sec: 3D superconductors as a topological phase}
and study their many-body wave functions and topological degeneracies.
We conclude in Sec.\ \ref{sec: discussions} with
a brief discussion of the close connection between the non-trivial 
topological characteristics of the 3D bulk and
the Anderson (de)localization physics at the 2D boundary.

For those readers who are interested in details and
prefer a systematic presentation,
we recommend to read all sections sequentially.
On the other hand, those readers who prefer to understand
the concepts rather through explicit examples, may skip
Secs.\ \ref{sec: symmetry classification of non-interacting Hamiltonians}
and \ref{sec: characterization at the boundary},
and should proceed to read only
Secs.\
\ref{sec: characterization in the bulk},
\ref{sec: 3D Dirac Hamiltonian},
\ref{sec: topological field theory description},
\ref{sec: 3D superconductors as a topological phase},
and
\ref{sec: discussions}.

%%%%%%%%%%%%%%%%%%%%%%%%%%%%%%%%%%%%%%%%%%%%%%%%%%%%%%%%%%%%%%%%%%%
\section{Symmetry classification of non-interacting Hamiltonians}
\label{sec: symmetry classification of non-interacting Hamiltonians}

We start by recalling the basic ideas underlying the 
classification by AZ \cite{Altland97,Bernard01}
of non-interacting fermionic Hamiltonians (``random matrices'')
in terms of ten symmetry classes.
These are classified in terms of the presence or absence
of certain discrete symmetries
(see Table \ref{tab: rmt} for a summary).
The corresponding symmetry operations are classified into two types,
\begin{subequations}
\label{DEFChiral}
\begin{align}
P:\quad
\mathcal{H} &= -P \mathcal{H} P^{-1},
\quad
PP^{\dag}=1,
\quad
P^2=1,
\\%%%%%
C:\quad
\mathcal{H} &= \epsilon_c C \mathcal{H}^T C^{-1},
\quad
CC^{\dag}=1,
\quad
C^T = \eta_c C,
\label{eq: def epsilon and eta}
\end{align}
\end{subequations}%
where $\mathcal{H}$ is a matrix or an operator
representing a single-particle Hamiltonian,
$\epsilon_c = \pm 1$
and
$\eta_c = \pm 1$.

The symmetry operation corresponding to $C$
($C$-type symmetry) represents
a TRS  operation when $\epsilon_c = 1$
and a PHS symmetry operation when $\epsilon_c = -1$.
Furthermore, we distinguish two cases,
$\eta_c = \pm 1$:
$(\epsilon_c,\eta_c)=(1,1)$ represents
a TRS for spinless (or integer spin) particles,
whereas $(\epsilon_c,\eta_c)=(1,-1)$ represents
a TRS for spinful, half-integer spin particles.
Similarly, 
$(\epsilon_c,\eta_c)=(-1,1)$ represents
a PHS for a triplet pairing BdG Hamiltonian
whereas $(\epsilon_c,\eta_c)=(-1,-1)$ represents
a PHS for a  singlet pairing BdG Hamiltonian.
Note that although the form of a $C$-type symmetry operation
can be changed by a unitary transformation,
the value of $(\epsilon_c,\eta_c)$ 
remains unchanged.
The presence of TRS for half-integer spin implies
Kramers degeneracy whereas the presence of PHS implies that
the energy spectrum is symmetric about zero energy.

Similar to PHS ($C$-type symmetry with $\epsilon_c=-1$),
a $P$-type symmetry implies a symmetry of the energy spectrum.
In condensed matter systems, it is often realized as a sublattice
symmetry on a bipartite lattice
(i.e., the symmetry operation that changes the sign of wave functions
on all sites of one of the two sublattices of the bipartite lattice)
and is sometimes called {\it chiral symmetry}.
When we have two $P$-type symmetries, $P$ and $P'$, say,
we can construct a conserved quantity by combining them,
i.e. 
$
\left[\mathcal{H}, PP'\right]
=0
$
where $PP'(PP')^{\dag}=1$.
Note that the latter property implies that we can block-diagonalize
$\mathcal{H}$ and
apply our classification
scheme to each of the blocks.
Consequently, it is enough to consider
ensembles of Hamiltonians
possessing only a single, or no $P$-type symmetry.

Note that whenever a Hamiltonian possesses both
$P$- and $C$-type symmetries,
it automatically has another, different $C$-type symmetry $C'$
defined by
\begin{align}
\mathcal{H} &=
\epsilon'_c C' \mathcal{H}^T C^{\prime -1},
\quad
C' = PC,
\quad
\epsilon'_c=-\epsilon^{\ }_c.
\end{align}
Thus, symmetry classes of Hamiltonians possessing both
$P$- and $C$-type symmetries automatically
possess in fact all three, chiral, particle-hole and
time-reversal symmetries.

The complete classification in terms of the presence or absence
of chiral, particle-hole, and time-reversal symmetries
is summarized in Table \ref{tab: rmt}.
The ten symmetry classes of AZ can be grouped into three categories:
the three standard (or ``Wigner-Dyson'') classes 
\{A, AI, AII\},
the three chiral classes 
\{AIII, BDI, CII\},
and the four BdG (superconductor) classes
\{D, C, DIII, CI\}.
We wish to point out, however, that classes CI and DIII can
be thought of as a close cousin of
the chiral classes, since in each of these two
classes one can find a unitary matrix,
by combining the TRS and PHS,
which anticommutes with all members of the class.
Conversely, class AIII can also be thought of as a  
BdG (superconductor) class.\cite{Foster07}
(This will be discussed further below).

Below, we will give a more
detailed and physical description for each class.
In this section, we use four sets of standard Pauli matrices
$s_{\mu}$, $c_{\mu}$, $t_{\mu}$, and $r_{\mu}$ 
(where $\mu=0,x,y,z$, and
$s_0$, $c_0$, $t_0$, and $r_0$ are $2\times 2$ unit matrices.)
Unless otherwise specified, the Pauli matrices
$s_{\mu}$ act on spin indices ($\uparrow$/$\downarrow$),
whereas
$c_{\mu}$ act on the two ($A$- and $B$-) sublattice indices
of a bipartite lattice;
the Pauli matrices
$t_{\mu}$ are used to represent
the particle-hole space appearing in the BdG Hamiltonian for
quasiparticles in a superconductor,
whereas $r_{\mu}$ are used for superconductors
for which the $z$-component $S_z$ of spin is conserved.
We also use two additional sets of Pauli matrices,
$\sigma_{\mu}$, $\tau_{\mu}$
(where $\mu=0,x,y,z$, 
and $\sigma_0$ and $\tau_0$
are $2\times 2$ unit matrices).

\subsection{Standard (Wigner-Dyson) classes}

Let us first review the familiar, standard
Wigner-Dyson symmetry classes.\cite{MehtaRandomMatrices}
An ensemble of Hamiltonians without any constraint
other than being Hermitian
is called the unitary symmetry class (class A).
Imposing TRS for half-integer spin
on the unitary symmetry class,
\begin{align}
{i}s_y \mathcal{H}^T (-{i}s_y)
&=
\mathcal{H},
\label{eq: TRS (odd)}
\end{align}
one obtains the symplectic symmetry class (class AII);
note that ${i}s_y$ is antisymmetric,
$({i}s_y)^T=-{i}s_y$.
Imposing, in addition, SU(2) spin rotation symmetry
on the symplectic symmetry class,
one obtains the orthogonal symmetry class (class AI),
\begin{align}
\mathcal{H}^T
&=
\mathcal{H}.
\label{eq: TRS (even)}
\end{align}
The symmetry operation in Eq.\ (\ref{eq: TRS (even)})
is the TRS operation for integer-spin or spinless particles.
We distinguish the two $C$-type symmetries in Eqs.\ (\ref{eq: TRS (odd)})
and (\ref{eq: TRS (even)}) by 
referring to them as ``TRS (odd)'' and ``TRS (even)'', respectively.

\subsection{Chiral classes}

Symmetry classes of Hamiltonians possessing
a $P$-type symmetry implemented by
a unitary transformation
\begin{eqnarray}
c_z \mathcal{H}c_z = -\mathcal{H},
\label{eq:chiral}
\end{eqnarray}
[letting $c_z := P$ in Eq.\ (\ref{DEFChiral})]
are conventionally called chiral classes.
As already mentioned above,
the corresponding unitary transformation is typically implemented
as a sublattice symmetry on a bipartite lattice.
Equations (\ref{DEFChiral}) and (\ref{eq:chiral}) imply
that all the energy eigenvalues appear in pairs
(with a possible exception at zero energy).
From an eigenstate $\psi$ with energy $E$,
one can obtain another state with the opposite
energy $-E$ by a unitary transformation, $c_z \psi$.

In complete analogy with the standard (``Wigner-Dyson'') classes
discussed above,
the ensemble of {\it chiral} Hamiltonians
without any further conditions is called
the chiral unitary class (class AIII).
Imposing TRS for half-integer spin (\ref{eq: TRS (odd)}),
we obtain the chiral symplectic class (class CII).
Imposing both the TRS and SU(2) symmetries
[i.e., imposing the TRS (\ref{eq: TRS (even)})],
we obtain the chiral orthogonal symmetry class (class BDI).

A well-known physical realization of
the chiral unitary symmetry class (class AIII)
is a disordered tight-binding model on
a bipartite lattice with broken TRS,
such as the random flux problem.\cite{Furusaki99}
However, a class AIII Hamiltonian can \cite{Foster07}
also be interpreted as an ensemble of
BdG Hamiltonians that have a TRS and are invariant
under a U(1) subgroup of SU(2) spin rotation symmetry
(rotation around $z$-component of spin, say),
as will be discussed further below.

\subsection{BdG classes}
\label{sec: symmetry classification of non-interacting Hamiltonians,BdG}

\begin{table*}[t]
\begin{center}
\begin{tabular}{|c||c|c||l||l|}\hline
AZ class  &  $\mathrm{SU}(2)$    & TRS &  $\,$Constraints on Hamiltonians & $\,$Examples in 2D
\\\hline\hline
D    & $\times$    &$\times$ &  $\,$$t_x \mathcal{H}^T t_x = -\mathcal{H}$   & $\,$Spinless chiral $(p\pm {i}p)$-wave \\ \hline
DIII & $\times$   &$\bigcirc$ &  $\,$$t_x \mathcal{H}^T t_x = -\mathcal{H}$,  $ {i} s_y \mathcal{H}^T (-{i}s_y)=\mathcal{H}$  &  $\,$Superposition of $(p+ {i}p)$- and $(p-{i}p)$-wave
\\ \hline \hline
A    &   $\triangle$   &$\times$  &  $\,$ no constraint  & $\,$Spinful chiral $(p\pm {i} p)$-wave \\ \hline
AIII  & $\triangle$   &$\bigcirc$  &  $\,$$r_y \mathcal{H} r_y = -\mathcal{H}$    &  $\,$Spinful $p_x$- or $p_y$-wave  \\ \hline \hline
C    &$\bigcirc$     &$\times$    & $\,$$r_y \mathcal{H}^T r_y = -\mathcal{H}$   &  $\,$$(d\pm {i}d)$-wave \\ \hline
CI   & $\bigcirc$    & $\bigcirc$ &  $\,$$r_y \mathcal{H}^T r_y = -\mathcal{H}$, $\mathcal{H}^*=\mathcal{H}$   & $\,$$d_{x^2-y^2}$- or $d_{xy}$-wave \\ \hline
\end{tabular}
\caption{
\label{tab: bdg}
Symmetry classification of Bogoliubov-de Gennes systems, 
in terms of the presence (``$\bigcirc$'') or absence (``$\times$'') of $\mathrm{SU}(2)$ spin rotation symmetry and 
time-reversal symmetry (TRS).
In classes A and AIII, 
Hamiltonians are 
invariant under rotations about the $z$- (or any fixed)
axis in spin space, but not under full $\mathrm{SU}(2)$ rotations,
as denoted by ``$\triangle$'' in the table.
The sets of standard Pauli matrices
$s_{x,y,z}$, $t_{x,y,z}$, and $r_{x,y,z}$ 
act on
the spin,
the particle-hole,
and 
the grading defined in Eq.\ (\ref{H_2}),
respectively.
}
\end{center}
\end{table*}

Following Altland and Zirnbauer,\cite{Altland97}
we now consider the following general form of a Bogoliubov-de Gennes Hamiltonian
for the dynamics of quasiparticles deep inside the superconducting
state of a superconductor
\begin{eqnarray}
H
=\frac{1}{2}
\left(
\begin{array}{cc}
\boldsymbol{c}^{\dag}, & \boldsymbol{c}
\end{array}
\right)
\mathcal{H}_4
\left(
\begin{array}{c}
\boldsymbol{c}^{\ } \\
\boldsymbol{c}^{\dag}
\end{array}
\right),\,\,
\mathcal{H}_4=
\left(
\begin{array}{cc}
\Xi & \Delta \\
 -\Delta^{*} & -\Xi^{{T}}
\end{array}
\right),
\label{eq: BdG hamiltonian}
\end{eqnarray}
where $\mathcal{H}_4$ is a $4N \times 4N$ matrix for a system with $N$
orbitals (lattice sites),
and $\boldsymbol{c}=
\left(\boldsymbol{c}_{\uparrow},\boldsymbol{c}_{\downarrow}\right)$.
[$\boldsymbol{c}$ and $\boldsymbol{c}^{\dag}$ 
can be either column or row vector
depending on the context.]
Because of
$\Xi=\Xi^{\dag}$ (hermiticity)
and
$\Delta=-\Delta^{{T}}$ (Fermi statistics),
the BdG Hamiltonian (\ref{eq: BdG hamiltonian})
satisfies
\begin{eqnarray}
(a):&&
\mathcal{H}^{\ }_4=-t_{x}\mathcal{H}^{{T}}_4 t_{x},
\quad
\mbox{[PHS (triplet)]}.
\label{eq: def PHS (triplet)}
\end{eqnarray}
This is a $C$-type symmetry
with $(\epsilon_c,\eta_c)=(-1, +1)$,
and will be called the PHS (triplet).

In terms of the presence or absence of TRS (odd) represented by
\begin{eqnarray}
(b):&&
\mathcal{H}^{\ }_4= {i}s_y \mathcal{H}^{{T}}_4 (-{i}s_y),
\quad
\mbox{[TRS (odd)]},
\end{eqnarray}
and of $\mathrm{SU}(2)$ spin rotation symmetry represented by
\begin{eqnarray}
(c):&&
\big[\mathcal{H}_4,J_{a}\big]=0,
\quad
J_{a}:=
\left(
\begin{array}{cc}
s_{a} & 0 \\
0 & -s_{a}^{{T}}
\end{array}
\right),
\nonumber \\%%%%%
&&
\quad
a=x,y,z,
\quad
\mbox{[SU(2) symmetry]},
\end{eqnarray}
BdG Hamiltonians (\ref{eq: BdG hamiltonian}) are classified into the
four sub classes listed in Table \ref{tab: rmt}:
Classes C and CI are primarily relevant to singlet SC whereas
classes D and DIII are primarily relevant to triplet SC,
although one can also consider admixture of singlet and triplet
order parameters in the absence of the parity symmetry,
as known in, e.g., $\mathrm{CePt}_3\mathrm{Si}$.
\cite{Bauer04}

\subsubsection{BdG classes without spin rotation symmetry}

\textbf{Class D}\,
First, we consider the symmetry class with neither
TRS nor $\mathrm{SU}(2)$ invariance.
In this case,
a set of BdG Hamiltonians satisfying $(a)$
is nothing but the Lie algebra $\mathrm{so}(4m)$.
Any element $\mathcal{H}_4 \in \mathrm{so}(4m)$
can be diagonalized by an
$\mathrm{SO}(4m)$ matrix $g$ as
$
g \mathcal{H}_4 g^{-1}
=\mathrm{diag}( \varepsilon,-\varepsilon),
$
with
$
\varepsilon=\mathrm{diag}(\varepsilon_1,\varepsilon_2,\cdots)
$,
i.e., the spectrum is particle-hole symmetric.

An example of class D BdG Hamiltonian in 2D is
a 2D spinless chiral $p$-wave ($p\pm {i}p$-wave) superconductor,
which can be written in momentum space as
\begin{eqnarray}
&&
H =
\frac{1}{2}
\sum_k
\left(
\begin{array}{cc}
c^{\dag}_k, & c^{\ }_{-k}
\end{array}
\right)
h(k)
\left(
\begin{array}{c}
c^{\ }_{k} \\
c^{\dag}_{-k}
\end{array}
\right),
\nonumber \\%%%%%
&&
h(k) =
\bar{\Delta} \left(k_x t_x + k_y t_y \right)
+ \varepsilon_k t_z,
\label{eq: chiral p-wave} 
\end{eqnarray}
where $k=(k_x,k_y)$ is the 2D momentum, 
$\bar{\Delta}\in \mathbb{R}$ is the amplitude of the order parameter,
and $\varepsilon_k$ denotes the energy dispersion of a single particle.
The Hamiltonian has PHS (triplet)
[Eq.\ (\ref{eq: def PHS (triplet)})],
$
h(k)=-t_x h^T(-k)t_x 
$.

\textbf{Class DIII}\,
Consider class DIII, which satisfies
conditions $(a)$ and $(b)$.
A set of matrices which simultaneously satisfy
$(a)$ and $(b)$ does not form a subalgebra of $\mathrm{so}(4m)$,
but consists of all those elements of the Lie algebra 
$\mathrm{so}(4m)$ which are not elements of the
sub Lie algebra $\mathrm{u}(2m)$.
\cite{Altland97}

Combining $(a)$ and $(b)$, one can see that
a member of class DIII anticommutes with the unitary matrix $t_x \otimes s_y$,
\begin{eqnarray}
\mathcal{H}^{\ }_4 =
-t_x \otimes s_y \mathcal{H}_4 t_x\otimes s_y.
\end{eqnarray}
In this sense, class DIII Hamiltonians have
a chiral structure. It is sometimes convenient
to take a basis in which the chiral transformation,
which is $t_x \otimes s_y$ in the present basis,
is diagonal. In one of such bases,
a class DIII Hamiltonian takes on the form
\begin{eqnarray}
\mathcal{H}_4 =
\left(
\begin{array}{cc}
0 & D \\
D^{\dag} & 0
\end{array}
\right),
\quad
D = -D^T .
\label{eq: chiral basis for DIII}
\end{eqnarray}

An example of a 2D BdG Hamiltonian in symmetry class DIII is 
a $p_{x}$-wave (or $p_{y}$-wave) superconductor
with $d$-vector not pointing the $z$ direction.
An equal superposition of two chiral $p$-wave SCs
with opposite chiralities ($p_x+{i}p_y$- and $p_x-{i}p_y$-wave)
\cite{RoySC,SenguptaRoySC2,RoySF2008,Xiao-LiangQiTopSC08}
also falls into this class.
The latter can be 
explicitly written in momentum space as
\begin{subequations} \label{eq: BdG triplet in k space}
\begin{eqnarray} \label{p_x+i p_y and p_x - ip_y}
H 
=
\frac{1}{2}
\sum_k
\left(
\begin{array}{cc}
\boldsymbol{c}^{\dag}_k, & 
\boldsymbol{c}^{\ }_{-k}
\end{array}
\right)
 \left(
 \begin{array}{cc}
\Xi^{\ }_k & \Delta^{\ }_k \\
\Delta^{\dag}_k & -\Xi^T_{-k}
 \end{array}
 \right)
\left(
\begin{array}{c}
\boldsymbol{c}^{\ }_k \\
\boldsymbol{c}^{\dag}_{-k}
\end{array}
\right),
\end{eqnarray}
with the row vector
$
(
\boldsymbol{c}^{\dag}_k, 
\boldsymbol{c}^{\ }_{-k}
)
=
(
  c^{\dag}_{k\uparrow}, c^{\dag}_{k\downarrow}, c^{\ }_{-k\uparrow}, c^{\ }_{-k\downarrow}
)$, and  the matrix elements
\begin{eqnarray}
\Xi_k \!\!&=&\!\!
\varepsilon_k s_0,
\nonumber \\%%%%%
\Delta_k 
\!\!&=&\!\! 
\bar{\Delta}
\left(
\begin{array}{cc}
k_x + {i}k_y & 0 \\
0 & -k_x + {i}k_y
\end{array}
\right) .
\label{eq: p+ip x p-ip SC}
\end{eqnarray}
In terms of the $d$-vector,
$
\boldsymbol{d}_k \!  =\!  
\bar{\Delta}
\left(
\begin{array}{ccc}
-k_x, & k_y, & 0
\end{array}
\right)
$,
the superconducting order parameter reads
\begin{eqnarray} \label{eq: BdG triplet in k space_c}
\Delta_{k}
=
(\boldsymbol{d}_k \cdot \bm{s}) ({i}s_y) .
\end{eqnarray}
\end{subequations}
It is interesting to note,
that Hamiltonian~(\ref{eq: BdG triplet in k space}) is 
a direct product of Hamiltonian~(\ref{eq: chiral p-wave}),
$h(k_x,k_y)$, and $h(-k_x,k_y)$.
This follows from a
simple reordering of the
basis elements in Eq.~(\ref{eq: BdG triplet in k space}), such that
$( \bm{c}^{\dag}_k, \bm{c}^{\ }_{-k} )
\to
(
  c^{\dag}_{k\uparrow}, c^{\ }_{-k\uparrow}, c^{\dag}_{k\downarrow}, c^{\ }_{-k\downarrow}
)$.
The superconductor described by
the order parameter Eq.\ (\ref{eq: p+ip x p-ip SC})
can be thought of as a two-dimensional analogue of
the BW state realized in the B phase of ${ }^{3}\mathrm{He}$.
The BW state,  which is also a member of class DIII
and described by the $d$-vector 
$\boldsymbol{d}_k=
\bar{\Delta}
\left(
\begin{array}{ccc}
k_x, & k_y, & k_z
\end{array}
\right)
$,
will be discussed in
Sec.\ \ref{sec: 3D Dirac Hamiltonian} 
as an example of 3D topological insulators.

\subsubsection{BdG classes with $S_z$ conservation}

Let us consider BdG Hamiltonians
which are invariant under rotations about the $z$- (or any fixed)
axis in spin space,
yielding to the condition
$\left[ \mathcal{H}_4, J_{z} \right]= 0$,
which implies that the Hamiltonian can be brought into the form
\begin{eqnarray}
\mathcal{H}_4=
\left(
\begin{array}{cccc}
a & 0 & 0& b \\
 0 & a' & -b^T & 0\\
0  & -b^{*} & -a^T & 0\\
b^{\dag} & 0&0 & -a^{\prime T}
\end{array}
\right)\!,
\,\,
a^{\dag}=a,
\,\,
a^{\prime \dag}=a^{\prime}.
\end{eqnarray}
Due to the sparse structure of $\mathcal{H}_4$,
we can rearrange the elements of this
$4N \times 4N$ matrix into the form of a $2N\times 2N$ matrix,
\begin{equation}
H
=
\left(
\begin{array}{cc}
\boldsymbol{c}_{\uparrow}^{\dag}, &
\boldsymbol{c}^{\ }_{\downarrow}
\end{array}
\right)
\left(
\begin{array}{cc}
 a & b \\
b^{\dag} & -a^{\prime T}
\end{array}
\right)
\left(
\begin{array}{c}
\boldsymbol{c}^{\ }_{\uparrow} \\
\boldsymbol{c}_{\downarrow}^{\dag} \\
\end{array}
\right)
+
\frac{1}{2}
\mathrm{tr}
\left[
a'-a
\right].
\end{equation}
Note that the Hamiltonian $H$ is traceless.

To summarize, a BdG Hamiltonian which is invariant under
rotations about the $z$-component in spin space can 
be brought
(up to a term which is proportional to the identity matrix)
into the form
\begin{align}
H
&=
\left(
\begin{array}{cc}
\boldsymbol{c}_{\uparrow}^{\dag}, &
\boldsymbol{c}^{\ }_{\downarrow}
\end{array}
\right)
\mathcal{H}_2
\left(
\begin{array}{c}
\boldsymbol{c}^{\ }_{\uparrow} \\
\boldsymbol{c}_{\downarrow}^{\dag} \\
\end{array}
\right),
\quad
\mathcal{H}_2
=
\left(
\begin{array}{cc}
\xi^{\ }_{\uparrow} & \delta \\
\delta^{\dag} & -\xi^{T}_{\downarrow}
\end{array}
\right),
\label{H_2}
\end{align}
where $\xi^{\dag}_{\sigma}=\xi^{\ }_{\sigma}$.
Without further constraints,
this Hamiltonian is a member of class A
(unitary symmetry class).
A physical realization of this class is
a 2D spinfull chiral $p$-wave ($p\pm {i}p$-wave) superconductor
[c.f. Eq.~(\ref{eq: BdG triplet in k space})]
with $d$-vector parallel to the $z$-direction,
\begin{eqnarray}
\boldsymbol{d}_k = 
\hat{z}
\bar{\Delta}
(k_x + {i}k_y)
=
\bar{\Delta}
\left(
\begin{array}{ccc}
0, & 0, & k_x + {i}k_y
\end{array}
\right).
\label{eq: spinful chiral p}
\end{eqnarray}
More compactly, the spinful chiral ($p\pm ip$)-wave
superconductor can be expressed as
\begin{eqnarray} 
H \!\!&=&\!\!
\sum_k 
\left(
\begin{array}{cc}
  c^{\dag}_{k\uparrow}, & c^{\ }_{-k\downarrow}
\end{array}
\right)
\left(
\begin{array}{cc}
\xi_{\uparrow k} & \delta^{\ }_k \\
\delta^{\dag}_k & -\xi_{\downarrow-k}
\end{array}
\right)
\left(
\begin{array}{c}
c^{\ }_{k\uparrow} \\
c^{\dag}_{-k\downarrow}
\end{array}
\right),
\label{eq: BdG singlet in k space}
\end{eqnarray}
where
\begin{eqnarray}
\delta_k
=
\bar{\Delta} (k_x + {i}k_y),
\label{eq: spinful chiral p-wave}
\end{eqnarray}
and 
$\xi^T_{\uparrow/\downarrow, k}=\xi^{\ }_{\uparrow/\downarrow, -k}$,
$\delta^T_{k}=\delta^{\ }_{-k}$
(by definition).
[Incidentally, the single-particle Hamiltonian defined by
Eqs.\ (\ref{eq: BdG singlet in k space})
and 
(\ref{eq: spinful chiral p-wave})
happens to belong to class D,
if $\xi_{\uparrow k}=\xi_{\downarrow k}$.
The Anderson-Brinkman-Morel (ABM) state,
which is 3D superfluid 
given
by the same order parameter as 2D spinful chiral $p$-wave 
superconductor [Eq.\ (\ref{eq: spinful chiral p})],
is also a member of class A.

\textbf{Class C (full $\mathrm{SU}(2)$ symmetry)}\,
If we further impose the full SU(2) rotation symmetry,
$\xi_{\sigma}$ and $\delta$ are constrained by
\begin{eqnarray}
\xi_{\downarrow}=\xi_{\uparrow}=:\xi,
\quad
\delta =\delta^T.
\label{eq: class C symmetry}
\end{eqnarray}
These two conditions can be summarized as
\begin{eqnarray}
r^{\ }_y \mathcal{H}^T_2 r^{\ }_y
=
-\mathcal{H}^{\ }_2,
\quad
\mbox{[PHS (singlet)]}.
\end{eqnarray}
This is a $C$-type symmetry
with $(\epsilon_c,\eta_c)=(-1, -1)$,
and will be called the PHS (singlet).

An example of a  BdG Hamiltonian in class C  in two dimensions is
the ($d+{i}d$)-wave superconductor.\cite{SenthilMarstonFisher}
Its Hamiltonian is given by Eq.~(\ref{eq: BdG singlet in k space}) together
with the matrix elements
$\xi_{ \uparrow k}
=
\xi_{ \downarrow k}
=
\varepsilon_k$ and
\begin{eqnarray}
\delta_k
\!\!&=&\!\!
\Delta_{x^2-y^2}
\left(
k^2_x - k^2_y
\right)
+
{i}
\Delta_{xy} k_x k_y,
\label{eq: d+id}
\end{eqnarray}
where $\Delta_{x^2-y^2}$ and  $\Delta_{xy} $
are real amplitudes for the $d_{x^2-y^2}$ and
${i}d_{xy}$ superconducting order parameters, respectively.

\textbf{Class CI (full $\mathrm{SU}(2)$ symmetry + TRS)}\,
Imposing both full SU(2) rotation and TR symmetries,
leads, 
in addition to the constraints (\ref{eq: class C symmetry}),
to
\begin{eqnarray}
\xi^* = \xi,
\quad
\delta^* =\delta.
\end{eqnarray}
These conditions can be summarized by
\begin{eqnarray}
r_y \mathcal{H}^T_2 r_y
=
-\mathcal{H}^{\ }_2,
\quad
\mathcal{H}^*_2 =
\mathcal{H}^{\ }_2.
\end{eqnarray}

Combining these two conditions
we can obtain a $P$-type (i.e., chiral) symmetry,
$r_y \mathcal{H}_2 r_y
=
- \mathcal{H}_2
$.
It is also convenient to rewrite the Hamiltonian
by rotating the $r_{\mu}$ matrices by
$(r_x,r_y,r_z)
\Rightarrow
(r_x,-r_z,r_y).
$
In this basis,
the class CI Hamiltonian takes
on block off-diagonal form \cite{Altland02}
\begin{eqnarray}
\mathcal{H}_2
=
\left(
\begin{array}{cc}
0 & D \\
D^{\dag} & 0
\end{array}
\right),
\quad
D=\delta-{i}\xi =D^T.
\label{eq: chiral rep for class CI}
\end{eqnarray}

An example of a BdG Hamiltonian in
class CI in two dimensions is
a 2D $d$-wave ($d_{x^2-y^2}$-wave)
superconductor,
which is described by the order parameter 
Eq.\ (\ref{eq: d+id}), with $\Delta_{xy}=0$, but with $\Delta_{x^2-y^2}\neq 0$.

\textbf{Class AIII ($S_z$ conservation + TRS)}\,
If,
in addition to conservation of the $z$-component
of spin we further impose TRS $(b)$,
we obtain the conditions
\begin{eqnarray}
\xi^T_{\uparrow}
=
\xi^{\ }_{\downarrow},\quad
\delta =\delta^{\dag},
\label{TRS for AIII}
\end{eqnarray}
which can be summarized as
\begin{eqnarray}
r_y \mathcal{H}_2 r_y
=
-
\mathcal{H}_2.
\label{eq: class AIII condition in bdg H2}
\end{eqnarray}
By interpreting the $r$-grading
as $c$-grading,
this is nothing but the SLS.
Thus, a BdG Hamiltonian possessing
TRS $(b)$ and
conserving one component of $\mathrm{SU}(2)$ spin,
can be thought of as a member of
the chiral class without TRS (AIII).\cite{Foster07}
An example of
a BdG Hamiltonian of a two-dimensional superconductor in
class AIII is a $p$-wave ($p_{x,y}$-wave) superconductor
with the $d$-vector parallel to the $z$-direction,
i.e., Hamiltonian~(\ref{eq: BdG triplet in k space}) with 
$\boldsymbol{d}_k = \bar{\Delta}\hat{z} k_{x,y}$
in Eq.\ (\ref{eq: BdG triplet in k space_c}),
or, alternatively, Hamiltonian (\ref{eq: BdG singlet in k space})
with
$\delta_{k} = \bar{\Delta} k_{x,y}$.

Before closing this section,
we emphasize that the symmetry classification we have described
also applies to interacting fermion systems,
as discrete symmetries can be imposed on 
the second quantized fermion creation and annihilation operators.

%%%%%%%%%%%%%%%%%%%%%%%%%%%%%%%%%%%%%%%%%%%%%%%%%%%%%%%%%%%%%%%%%%%
\section{Characterization in the bulk}
\label{sec: characterization in the bulk}

A useful quantity to discuss the bulk
characteristics of topological insulators
is the spectral projection operator. 
We will first discuss the spectral projector
for a general Bloch Hamiltonian with a bulk band gap,
and then specialize to those symmetry classes 
that satisfy a certain discrete unitary symmetry
(called $P$-type in 
Sec.\ \ref{sec: symmetry classification of non-interacting Hamiltonians}),
in which case the projection operator can be
brought into block off-diagonal form.
This block off-diagonal representation
of the projector allows for the definition
of a winding number that distinguishes between
different topological phases.

\subsection{Projection operator}

In the presence of translation invariance,
ground states of non-interacting fermion systems
can be constructed as a filled Fermi sea
in the $d$-dimensional Brillouin zone (BZ),
in Fourier space.
The band structure can then be viewed as a
map from the BZ to the space of Bloch Hamiltonians.
Similarly, the spectral projection operator~\cite{Avron83}
can be thought of as a map from the reciprocal 
unit cell to a certain Lie group or coset manifold,
which we will call space of projectors or target space.
In order to define the spectral projector,
let us consider an eigenvalue problem at momentum $k$,
\begin{eqnarray}
\mathcal{H}(k)|u_{\hat{a}}(k)\rangle
=
E_{\hat{a}}(k)
|u_{\hat{a}} (k)\rangle ,
\end{eqnarray}
where
$\mathcal{H}(k)$ is a Hamiltonian at $k$ in the BZ,
$|u_{\hat{a}} (k)\rangle$ 
is the Bloch state in the $\hat{a}$-th band 
with energy $E_{\hat{a}}(k)$.
We assume the existence of a bulk band gap
centered around some energy $E_0$ and
define the quantum ground state  
by filling all states with $E_{\hat{a}}(k)< E_0$ .
Without loss of generality, we can set $E_0 =0$,
as a suitable constant
chemical potential
in the definition of single-particle energy levels
can always be included.
The projector onto the filled Bloch states
at fixed $k$ is then defined as
\begin{eqnarray}
P (k) =
\sum_{\hat{a}}^{\mathrm{filled}}
|u_{\hat{a}}(k)\rangle
\langle u_{\hat{a}}(k)|.
\end{eqnarray}

It is convenient to introduce   
\begin{eqnarray}
Q(k)= 2P(k) -1.
\end{eqnarray}
It is readily checked that the so-defined
``$Q$-matrix'' is characterized by the conditions
\begin{eqnarray}
&&
Q^{\dag}=Q,
\quad
Q^2=1,
\quad
\mathrm{tr}\,Q = m-n,
\end{eqnarray}
where we consider the situation where, at each $k$,
we have $m$ filled and $n$ empty Bloch states.
Depending on the symmetry class,
additional conditions may be imposed on $Q$,
which we will consider later.
Without any such further conditions,
the projector takes values in the
so-called Grassmannian $G_{m,m+n}(\mathbb{C})$:
the set of eigenvectors
can be thought of as a unitary matrix, a member of
$\mathrm{U}(m+n)$.
Once we consider a projection onto the occupied states,
we have a gauge symmetry $\mathrm{U}(m)$ for
 the occupied states, and
a similar gauge symmetry
$\mathrm{U}(n)$ for the empty states.
Thus,
each projector is described by an element of the coset
$\mathrm{U}(m+n)/\mathrm{U}(m)\times \mathrm{U}(n)
\simeq
G_{m,m+n}(\mathbb{C})
\simeq
G_{n,m+n}(\mathbb{C})
$.
On the other hand,
an element of $G_{m,m+n}(\mathbb{C})$ can be written as
\begin{eqnarray}
Q=
U \Lambda U^{\dag},
\,\,
\Lambda =
\mathrm{diag}(\mathbb{I}_{m}, -\mathbb{I}_{n}),
\,\,
U\in \mathrm{U}(m+n).
\end{eqnarray}
(where $m$ eigenvalues
of the diagonal matrix $\Lambda$
equal $+1$, and the remaining
$n$ eigenvalues equal $-1$.)
Imposing TRS or PHS, which are realized by an anti-unitary operation 
(called $C$-type in 
Sec.\ \ref{sec: symmetry classification of non-interacting Hamiltonians}),
prohibits certain types of maps from the BZ to the space of projectors (see Table~\ref{tab: nlsm}).
On the other hand, if the discrete symmetry is realized by a unitary operation
(called $P$-type in 
Sec.\ \ref{sec: symmetry classification of non-interacting Hamiltonians}) 
the space of projectors is altered from $G_{n,m+n}(\mathbb{C})$
to $\mathrm{U}(m)$ (see the following subsection).

\begin{table*}[t]
\begin{center}
\begin{tabular}{|c||l||c|c||c|c|}\hline
AZ 
& $\,$ space of projectors in momentum space
& BL 
& $N^{\mathrm{min}}_f$
& fermionic replica
& topological or 
\\
class
&
& class
&
& NL$\sigma$M target space
& WZW term

           \\ \hline \hline 
A &
$\{\, Q(k)\in G_{m,m+n}(\mathbb{C})\, \}$
&
\textbf{0}
&
1
&
$\mathrm{U}(2N)/\mathrm{U}(N)\times \mathrm{U}(N)$ & Pruisken
\\ \hline
AI &
$\{\, Q(k)\in G_{m,m+n}(\mathbb{C})\, |\, Q(k)^*=Q(-k)\, \}$ &
$\textbf{4}_+$
&
2
&
$\mathrm{Sp}(2N)/\mathrm{Sp}(N)\times \mathrm{Sp}(N)$ &
 N/A
\\ \hline
AII &
$\{\, Q(k)\in G_{2m,2(m+n)}(\mathbb{C})\,|\,({i}\sigma_y)
 Q(k)^*(-{i}\sigma_y) =Q(-k)\, \}$
&
$\textbf{3}_+$
&
1
   & $\mathrm{O}(2N)/\mathrm{O}(N)\times \mathrm{O}(N)$ &
 $\mathbb{Z}_2$
\\ \hline \hline
AIII
& $\{\, q(k)\in \mathrm{U}(m)\, \}$
&
\textbf{1} or \textbf{2}
&
1 or 2
& $\mathrm{U}(N)\times \mathrm{U}(N)/ \mathrm{U}(N)$ &
 WZW
\\ \hline
BDI
&
$\{\, q(k)\in \mathrm{U}(m)|q(k)^*=q(-k)\, \}$
&
$\textbf{9}_+$
&
2
& $\mathrm{U}(2N)/ \mathrm{Sp}(N)$ & N/A
\\ \hline
CII
&
$\{\, q(k)\in \mathrm{U}(2m)\, |\,({i}\sigma_y) q(k)^*(-{i}\sigma_y)
 =q(-k)\, \}$
&
$\textbf{9}_-$
&
2
& $\mathrm{U}(2N)/ \mathrm{O}(2N)$ & $\mathbb{Z}_2$
\\ \hline \hline
D &
$\{\, Q(k)\in G_{m,2m}(\mathbb{C})\, |\, \tau_x Q(k)^* \tau_x = -Q(-k)\,\}$ &
$\textbf{3}_-$
&
1
   & $\mathrm{O}(2N)/ \mathrm{U}(N)$ & Pruisken
\\ \hline
C &
$\{\, Q(k)\in G_{m,2m}(\mathbb{C})\,|\,\tau_y Q(k)^* \tau_y = -Q(-k)\,\}$
&
$\textbf{4}_-$
&
2
  & $\mathrm{Sp}(N)/ \mathrm{U}(N)$ & Pruisken
\\ \hline
DIII &
$\{\, q(k)\in \mathrm{U}(2m)\,|\, q(k)^T = -q(-k)\,\}$
&
\textbf{5} or \textbf{7}
&
1 or 2
& 
$\mathrm{O}(2N)\times \mathrm{O}(2N)/\mathrm{O}(2N)$ & WZW
\\ \hline
CI
  &$\{\, q(k)\in \mathrm{U}(m) \, |\, q(k)^T =q(-k) \, \}$
&
\textbf{6} or \textbf{8}
&
2 or 4
  & $\mathrm{Sp}(N)\times \mathrm{Sp}(N)/\mathrm{Sp}(N)$ & WZW
\\ \hline
\end{tabular}
\caption{
\label{tab: nlsm}
The space of projectors in momentum space for each 
Altland-Zirnbauer (AZ) class.
The BL classes  represent the classification of 2D {\it  Dirac Hamiltonians}
obtained by Bernard and LeClair,\cite{Bernard01}
and $N^{\mathrm{min}}_f$ is the smallest possible number of
flavors of 2D two-component Dirac fermions.
The fermionic replica NL$\sigma$M target spaces,
with possible 2D critical behavior
[in terms of whether it is possible for a given NL$\sigma$M
to have a topological term, 
either of Pruisken (IQHE) type or $\mathbb{Z}_2$ type,
or a Wess-Zumino-Witten (WZW) term]
are also listed according to Refs.\
\onlinecite{Fendley00}
and
\onlinecite{Fendley01}.
}
\end{center}
\end{table*}

We now ask if any element of the set of
projectors within a given symmetry class 
can be continuously deformed into any other,
without closing the energy gap.
Mathematically, this is related to the
homotopy group of the (topological) space of projectors.
For two spatial dimensions the relevant homotopy group is
$\pi_2[G_{m,m+n}(\mathbb{C})] = \mathbb{Z}$,
implying that the projectors are classified by an integer (Chern number);
\cite{Thouless82,Kohmoto,Avron83}
projectors with different Chern numbers 
cannot be deformed into each other adiabatically;
this is the mathematical reason why there 
exists a series of distinct
2D quantum Hall insulators labeled by an integer,
which is the Hall conductivity $\sigma_{xy}$
(measured in natural units of $e^2/h$).
On the other hand, 
in three spatial dimensions and in the absence of additional discrete symmetries,
the relevant homotopy group is\cite{Novikov,Nakahara03}
\begin{eqnarray}
\pi_3[G_{m,m+n}(\mathbb{C})] \simeq \{e\},
\end{eqnarray}
where $\{e \}$ represents a group with only one element
(identity).
Thus, there is no notion of winding in 3D in this case.
\cite{Comment 3DIQHE}
[When $m=1$, there is an accidental winding,
$\pi_3[G_{1,2}(\mathbb{C})] = \mathbb{Z}$,
because of the Hopf map.
\cite{Nakahara03,Moore2008}]

This is not to say that there is no topological distinction
when additional discrete symmetries are imposed
on the projector.
For example, in class AII (the symplectic class),
the projector must also satisfy the condition
$({i}\sigma_y) Q(k)^T (-{i}\sigma_y) =Q(-k)$,
which arises from the presence of TRS.
Due to this additional constraint,
two different $Q$-field configurations
might not be continuously deformable
into one another.
Properties of the projector for each class are summarized
in Table \ref{tab: nlsm}.

\subsection{Block off-diagonal projection operators}

A symmetry realized by an anti-unitary operation
(i.e., TRS or PHS) relates
the projector at
wavevector $k$ and the one at 
wavevector $-k$.
Thus, the role of PHS or TRS is to
prohibit certain $Q$-field configurations in momentum space,
or to change the topology of the BZ by orbifolding $k\equiv -k$.

In contrast, imposition of a discrete symmetry which is
realized by a unitary operation
(i.e., SLS or a product of PHS and TRS)
imposes a condition on the projector at each $k$.
Thus, the role of this type of symmetry is to change
the target manifold of the projector.
The main focus below is on the projector
in those symmetry classes which possess (in some basis)
a block off-diagonal representation
of the Hamiltonian and of the projector,
due to the presence of a chiral symmetry.
This is the case for all three chiral classes,
AIII, BDI, and CII,
and for two of the four BdG classes,
classes CI and DIII.
For these symmetry classes, the $Q$-matrix 
can be brought (upon basis change)
into block off-diagonal form
\begin{eqnarray}
Q =
\left(
\begin{array}{cc}
0 & q \\
q^{\dag} & 0
\end{array}
\right).
\label{eq: off-diagonal projectors}
\end{eqnarray}
Since $Q^2=1$, one has $qq^{\dag}=q^{\dag}q=1$,
and thus $q$ is a member of $\mathrm{U}(m)$.
As before, one should bear in mind
that the $q$-matrix can further be subject to
several additional constraints coming from 
additional discrete symmetries
imposed on Hamiltonians.
Properties of the projector for each class are summarized
in Table \ref{tab: nlsm}.

The relevant homotopy group
for projectors that take the form
Eq.\ (\ref{eq: off-diagonal projectors})
is \cite{Novikov,Nakahara03}
\begin{eqnarray}
\pi_d\left[\mathrm{U}(m)\right]
\simeq
\left\{
\begin{array}{cc}
\{e \}, & \mbox{for $d$ even,} \\
\\
\mathbb{Z}, & \mbox{for $d$ odd,}
\end{array}
\right. 
\end{eqnarray}
for $m \ge (d+1)/2$,
instead of
$\pi_d[G_{m,m+n}(\mathbb{C})]$.

If a symmetry class of interest does
allow topologically non-trivial configuration of the projector,
a useful tool to investigate if a given quantum ground state belongs to
a non-trivial topological sector (in the space of projectors),
is a \textit{quantized} invariant.
In the IQHE, it is the quantized Hall conductivity $\sigma_{xy}$,
which is essentially
a winding number characteristic of
$\pi_2[G_{m,m+n}(\mathbb{C})] = \mathbb{Z}$.
In the $\mathbb{Z}_2$ topological insulators,
it is the $\mathbb{Z}_2$ invariant, 
\cite{KaneMele, Roy06, Moore06, Fu06_3Da,Fu06_3Db, SSLee07}
which can be constructed from 
the SU(2) Wilson loops that are quantized because of the TRS.
Below, we will introduce a topological invariant that is 
applicable for symmetry classes with 
block off-diagonal projector.

\subsection{Winding number in 3D}

Since $\pi_3[\mathrm{U}(m)] \simeq \mathbb{Z}$ (for $m \ge 2$),
there are, in three spatial dimensions,
topologically non-trivial configurations
in the space of projectors 
for those symmetry classes
for which the $Q$-matrix can
be brought into block off-diagonal form.
To characterize these distinct classes, we can define the winding number
\begin{equation}
\nu[q]
=
\int
\frac{d^3 k\,}{24\pi^2}
\epsilon^{\mu\nu\rho}\,
\mathrm{tr}\left[
\left(
q^{-1}\partial_{\mu}q \right)
\left(
q^{-1}\partial_{\nu}q \right)
\left(
q^{-1}\partial_{\rho}q
\right)
\right],
\label{eq: def winding number}
\end{equation}
where
$q(k)\in \mathrm{U}(m)$,
$\mu,\nu,\rho=k_x,k_y,k_z$,
and the integral extends 
over the entire
Brillouin zone  for lattice systems,
which is the three-torus $T^3$,
whereas for continuum models 
the domain of integration
in Eq.\ (\ref{eq: def winding number})
is topologically equivalent to three-sphere $S^3$.

In class AIII, the winding number $\nu[q]$ can be any integer.
Due to additional constraints on
the $q$-field in classes DIII, CI, BDI and CII,
all integers might not be realized.
One way to determine which integer values 
can be realized as 
winding numbers  in each symmetry class,
is to count the number of flavors 
of massless Dirac fermions allowed by the symmetries
at the surface. This counting will be done in the
next section, yielding the following results.

An arbitrary number of flavors can be realized
in symmetry classes AIII and DIII.
The gapless nature of the surface Dirac (Majorana) fermions
is, irrespective of the number of flavors,
stable against
any perturbations respecting the symmetries of
a given class, including disorder potentials.
\cite{comment Roy and Sengupta}

On the other hand,
only an  even number of flavors is allowed
in classes CI, BDI and CII.
This suggests that the winding number for
classes CI, BDI and CII can 
take on only even
integer values.
Indeed, in Sec.\ \ref{sec: 3D Dirac Hamiltonian},
we will construct an
 explicit example of a
topological insulator in class CI
in three spatial dimensions, 
with winding number $\nu=2$
and with two flavors of surface Dirac fermions.

While an even number of flavors of surface Dirac fermions
might appear, at first sight, to be possible in class BDI,
a detailed study of the form of generic
perturbations reveals that the gapless nature of
Dirac fermions at the surface is not protected in 
this symmetry class.
This suggests that the space of all Hamiltonians in class BDI
has no non-trivial topology. 
[See also the discussion around Eq.\ (\ref{eq: gapped 2D Dirac}).]

The stability against perturbations
of the gapless 2D surface Dirac fermions in class CII
depends on whether the number of flavors is an even or an odd
multiple of two: if it is even (odd) multiple,
the gapless spectrum is unstable (stable).
We thus expect that the space of all Hamiltonians in class CII
has a $\mathbb{Z}_2$ classification, 
as in class AII (the symplectic symmetry class).
Furthermore, this $\mathbb{Z}_2$ classification
has nothing to do with the winding number, 
as we will demonstrate in Sec.\ \ref{sec: 3D Dirac Hamiltonian} 
where we construct a 3D class CII insulator
with two flavors of surface Dirac fermions
yet with vanishing winding number.

Before closing this section,
several comments are in order.
(i)
Similar discussions are possible
in all odd 
spatial dimensions, where
a winding number can always be defined for
classes AIII, BDI, CII, CI and DIII.
In particular,
since $\pi_1[\mathrm{U}(m)] \simeq \mathbb{Z}$,
topologically non-trivial insulators 
characterized by an integer invariant 
can exist
in one spatial dimension (1D)
when there is a SLS.\cite{Ryu01}
On the other hand, 
in the presence of PHS,
\cite{Xiao-LiangQi08,Kitaev00,Kitaev05} 
or TRS (odd), 
\cite{Fu06}
the non-trivial topological features
in the bulk of 1D topological insulators can 
be characterized by U(1) or SU(2) Wilson loops
of the Berry connection, respectively.
Although the quantized values of the Wilson loops are not affected
by the choice of Bloch wave functions, as different
Bloch wave functions are related by a gauge transformation,
the quantized values of the Wilson loops do depend on 
the choice of the unit cell.
\cite{Fu06, RyuEE06}
When there is translation invariance, 
the choice of unit cell is arbitrary, 
whereas if we introduce a boundary, the choice of 
unit cell should be consistent with
the location of the boundary.  
It is in this sense, that
the (quantized) values of Wilson loops in 1D reflect
the boundary physics, and are
not solely determined from the bulk properties.
(ii)
The relevant homotopy group governing
the existence
of topological insulators
in two spatial dimensions is
$\pi_2[\mathrm{U}(m)]=\{e\}$.
This immediately tells us that
there are no topological insulators
in class AIII in 2D.
However, due to constraints arising from the
presence of additional discrete symmetries
there is still the possibility of having
2D topological phases in other ``chiral'' classes.
Indeed, in class DIII, for example, one can construct
a topological insulator from
the mixture of $p+{i}p$ and
$p-{i}p$ pairing states.
\cite{RoySC, SenguptaRoySC2,RoySF2008,Xiao-LiangQiTopSC08}
This state is a direct analogue of the Kane-Mele model
\cite{KaneMele}
on the honeycomb lattice,
which is the mixture of the two
Haldane models of the IQHE.
\cite{Haldane88}
(iii)
Finally,
the winding number defined 
above in momentum space
(which requires translational invariance)
can also be defined for disordered systems,
in a similar fashion in which the Chern number can be
defined for disordered systems.
\cite{Niu85}

%%%%%%%%%%%%%%%%%%%%%%%%%%%%%%%%%%%%%%%%%%%%%%%%%%%%%%%%%%%%%%%%%%%
\section{Characterization at the boundary}
\label{sec: characterization at the boundary}

A physical consequence of the non-trivial
topological properties of the quantum state in the bulk
is the appearance of the gapless boundary modes.
(Some explicit examples will be constructed in the next section.)
Conversely, most of the possible bulk phases in (3+1) dimensions
can be inferred by studying
their possible (2+1)-dimensional boundary physics.
In this section we consider,
following Bernard and LeClair,\cite{Bernard01}
the symmetry classification
of 2D Dirac Hamiltonians of the form
\begin{eqnarray}
\label{GeneralDiracHamiltonian}
\mathcal{H} =
\left(
\begin{array}{cc}
V_+ + V_- & -{i}\bar{\partial} + A_+ \\
-{i}\partial + A_- & V_+ - V_-
\end{array}
\right),
\end{eqnarray}
where
$V_{\pm}^{\ } = V_{\pm}^{\dag}$
and
$A^{\dag}_+=A^{\ }_-$
and
$
\partial = \partial_x - {i}\partial_y
$,
$
\bar{\partial} =
\partial_x + {i}\partial_y
$.
Possible dimensionalities
of the matrices $V_{\pm}$ and $A_{\pm}$
depend on the symmetry class as we will see below.

As before, we impose two types of discrete symmetries, $P$ and $C$.
The form of the matrices $P$ and $C$,
acting by conjugation, is constrained by
the requirement that they do not change the kinetic term
in Eq.\ (\ref{GeneralDiracHamiltonian}),
resulting in the following block diagonal form
\begin{eqnarray}
P =
\left(
\begin{array}{cc}
\gamma & 0 \\
0 & -\gamma
\end{array}
\right),
\quad
C =
\left(
\begin{array}{cc}
0 & \sigma \\
-\epsilon_c \sigma & 0
\end{array}
\right),
\end{eqnarray}
where $\gamma$ and $\sigma$ are a matrix satisfying
\begin{eqnarray}
\gamma\gamma^{\dag}=1,
\quad
\gamma^2 =1,
\quad
\sigma\sigma^{\dag}=1,
\quad
\sigma^T = -\eta_c \epsilon_c \sigma.
\end{eqnarray}

All possible forms of $\gamma$ and $\sigma$ are listed
in Ref.\ \onlinecite{Bernard01}.
Due to the Dirac kinetic term,
the Bernard-LeClair (BL) classification is finer
than the ten symmetry classes of AZ:
there are 13 symmetry classes denoted by
$\textbf{0}$,
$\textbf{1}$,
$\textbf{2}$,
$\textbf{3}_{\pm}$,
$\textbf{4}_{\pm}$,
$\textbf{5}$,
$\textbf{6}$,
$\textbf{7}$,
$\textbf{8}$
and
$\textbf{9}_{\pm}$.
While all AZ symmetry classes except AIII, DIII and CI
are in one-to-one correspondence with the BL classes,
\textit{two} BL classes correspond to each of
the AZ symmetry classes AIII, DIII and CI: 
BL classes \textbf{1} and \textbf{2} correspond to AIII,
BL classes \textbf{5} and \textbf{7} to DIII,
whereas
BL classes \textbf{6} and \textbf{8} correspond to CI.
(For a summary, see Table \ \ref{tab: nlsm}.)

Of direct relevance to our discussion
of 3D topological insulators is
the minimal number $N^{\mathrm{min}}_f$ of flavors 
in the BL classification.
For BL class $\textbf{3}_+$
(class AII in the AZ classification),
$N^{\mathrm{min}}_f$ is one.
Since a single flavor of 2D gapless Dirac fermion
cannot be realized on a 2D lattice
without breaking TRS,
the case with $N^{\mathrm{min}}_f=1$
(or $N^{\mathrm{min}}_f=\mbox{odd}$)
in class $\textbf{3}_+$ should correspond to
a state appearing at the two-dimensional surface of
a three-dimensional $\mathbb{Z}_2$ topological insulator in class AII.
This situation is indeed realized in the model of Fu-Kane-Mele.
\cite{Fu06_3Da,Fu06_3Db}
For BL classes
\textbf{1},
\textbf{5}
and
\textbf{6},
the minimal number $N^{\mathrm{min}}_f$
of Dirac fermion flavors is half
the minimal number of flavors
required for the BL classes
\textbf{2},
\textbf{7}
and
\textbf{8}, respectively.
(Compare e.g.\ Table \ \ref{tab: nlsm}.)
We thus expect that 2D Dirac fermion modes
in the BL classes
\textbf{1},
\textbf{5}
and
\textbf{6}
are realized as boundary states
of a {\it  non-trivial}  3D topological insulator
in classes AIII, DIII, and CI,
respectively, 
whereas those in BL classes
\textbf{2},
\textbf{7}
and
\textbf{8}
can be realized either directly on a 2D lattice
or, at a surface of topologically  trivial 3D insulators.
Finally
we will show in the next section 
by constructing an explicit example,
that for the BL symmetry class $\textbf{9}_-$ (CII),
the case with minimal flavors $N^{\mathrm{min}}_f=2$
can be realized as a surface state of a 3D topological insulator.

We will now argue that
the gapless nature of
2D Dirac Hamiltonians that can be realized
at a boundary of a topologically non-trivial 3D insulator
is stable against perturbations
$V_{\pm}$ and $A_{\pm}$.

We first look at
the BL classes
$\textbf{1}$, $\textbf{5}$, and
$\textbf{6}$
which correspond to
AIII, DIII, and CI,
respectively.
These classes
are special in that
the potentials
$V_{\pm}$ are not allowed\cite{Bernard01}
by the symmetries in these classes,
so that
\begin{eqnarray}
\mathcal{H} =
\left(
\begin{array}{cc}
0 & -{i}\bar{\partial}+A_+ \\
-{i}\partial+A_- & 0 \\
\end{array}
\right).
\end{eqnarray}
Since the only allowed perturbations are
of gauge type, one would expect that
these perturbations do not spoil the
gaplessness of the free
Dirac spectrum.
To see this, let us now try to find zero-energy modes.
We thus look for the solution
\begin{eqnarray}
\left(
\begin{array}{cc}
0 & k_+ + A_+ \\
k_- + A_- & 0\\
\end{array}
\right)
\left(
\begin{array}{c}
\chi_A \\
\chi_B
\end{array}
\right)
=
0,
\end{eqnarray}
where $k_{\pm}=k_x \pm {i}k_y$.
We assume that $A_{\pm}$ have been brought into diagonal
form by gauge transformations
$g_{\pm}$,
so that
$g^{\ }_{\pm}
A_{\pm}
g^{-1}_{\pm}
=
\Lambda_{\pm}$,
where
$
\Lambda^{\ }_{+}=\Lambda^*_-$,
$g^{\dag}_+ = g^{-1}_-$,
and
$\Lambda_{\pm}$ are diagonal matrices
with complex entries.
Thus, the Schr\"odinger equation 
for the zero modes reduces to
$
(
k_+ + \Lambda^{\ }_+
)\tilde{\chi}_B
=
(
k_- + \Lambda^*_+
)\tilde{\chi}_A = 0,
$
where
$\tilde{\chi}_B = g^{-1}_+ \chi_B$
and
$\tilde{\chi}_A = g^{-1}_- \chi_A$.
This wave equation has a non-trivial solution
($\tilde{\chi}_{A}\neq 0$
or $\tilde{\chi}_{B}\neq 0$),
only when one of complex eigenvalues of
$\Lambda_+$ is equal to $-k_+$.
The solution $\chi_A$ and $\chi_B$ can 
then be obtained
from $\tilde{\chi}_A$ and $\tilde{\chi}_B$.
It should be normalized as
$
\chi^{\dag}_A \chi^{\ }_A
+
\chi^{\dag}_B \chi^{\ }_B
=
\tilde{\chi}^{\dag}_A g^{\dag}_- g_-
\tilde{\chi}^{\ }_B
+
\tilde{\chi}^{\dag}_B g^{\dag}_+ g_+
\tilde{\chi}^{\ }_B
=1.
$
(Note that $g^{\dag}_+ \neq g^{-1}_+$).
Thus we conclude that
the gaplessness of the spectrum
of the 2D Dirac modes
appearing at the surface
of 3D topological insulators
in AZ symmetry classes
AIII, DIII, and CI,
are robust against arbitrary static perturbations.  
The location of the Dirac cones, however, 
might be shifted by the gauge type perturbations.
[As we will discuss,
the gaplessness
of the same Dirac surface modes is
also robust against Anderson localization  arising from
random perturbations (which break translational invariance),
because of the presence of Wess-Zumino-Witten (WZW) terms.]

We note that, on the other hand, it is also easy to see that for
all other classes except AII and CII,
the 2D surface modes have in general a massive spectrum.
To see this, take $V_+=A_+=A_-=0$ and consider the Hamiltonian
\begin{align}
\mathcal{H}
&=
k_x \sigma_x +
k_y \sigma_y +
V_- \sigma_z.
\label{eq: gapped 2D Dirac}
\end{align}
The potential $V_-$ can be diagonalized by a unitary matrix $U$,
$V_- \to
U^{\dag}V_- U
=:\Upsilon=\mathrm{diag}(\upsilon_n)
$,
$
\mathcal{H}
\to
k_x \sigma_x +
k_y \sigma_y +
\Upsilon \sigma_z
$,
and hence the eigenvalues are
$
E_n(k_x, k_y) = \pm \sqrt{k^2_x+k^2_y+\upsilon_n^2}
$.

In contrast, for classes AII and CII,
the potential $V_-$ satisfies
\begin{eqnarray}
V^{\ }_- = - V^T_- .
\end{eqnarray}
This guarantees that
the corresponding Dirac Hamiltonian
has at least one zero eigenvalue when
the dimensionality of $V_-$ is odd.
We thus cannot completely
gap out the spectrum by $V_-$.

%%%%%%%%%%%%%%%%%%%%%%%%%%%%%%%%%%%%%%%%%%%%%%%%%%%%%%%%%%%%%%%%%%%
\section{3D Dirac Hamiltonians}
\label{sec: 3D Dirac Hamiltonian}

The purpose of this section is to construct
examples of 3D topological insulators in the continuum,
one for each of the five classes AII, DIII, AIII, 
CI and CII. 
The examples we give are continuum 3D Dirac Hamiltonians
perturbed by a mass term of some sort.
For these examples we will compute the winding number $\nu$
introduced in Eq.\ (\ref{eq: def winding number})
and
explicitly derive the surface Dirac modes,
thereby illustrating
the above-mentioned connection between the 
topological properties of the bulk
and the existence of stable, massless
surface states. 
The Dirac Hamiltonians with the minimal number
of components have four components
for classes AII, DIII and AIII, 
whereas the minimal number of components is eight
for classes CI and CII.

\subsection{3D four-component Dirac Hamiltonian}
\label{sec: 3D four-component Dirac Hamiltonian}

\subsubsection{Hamiltonian and its symmetries}

Let us consider the following four-component (3+1)D massive Dirac Hamiltonian
\begin{eqnarray}
\mathcal{H} =
-{i}\partial_{\mu} \alpha_{\mu}
+
m \beta,
\quad
\mu=x,y,z,
\label{eq: 3D Dirac}
\end{eqnarray}
where $m\in \mathbb{R}$,
and we use the standard (or Dirac) representation of
the (3+1)D gamma matrices,
\begin{eqnarray}
\alpha_{\mu} \!\!&=&\!\!
\tau_x \otimes \sigma_{\mu}
=
\left(
\begin{array}{cc}
0 & \sigma_{\mu} \\
\sigma_{\mu} & 0
\end{array}
\right),
\quad
\beta =
\tau_z
=
\left(
\begin{array}{cc}
1 & 0 \\
0 & -1
\end{array}
\right),
\nonumber \\%%%%%
\gamma^5
\!\!&=&\!\!
\tau_x
=
\left(
\begin{array}{cc}
0 & 1\\
1 & 0
\end{array}
\right),
\quad
\mu=x,y,z.
\end{eqnarray}
In momentum space,
\begin{equation}
\mathcal{H}(k)
=
\alpha_{\mu}k_{\mu}
+
m\beta
=
\left(
\begin{array}{cc}
m & k\cdot \sigma\\
k\cdot \sigma & -m
\end{array}
\right),
\label{eq: 3D Dirac in k}
\end{equation}
and the energy spectrum is given
by $E(k)=\pm \sqrt{k^2+m^2}=:\pm \lambda(k)$
(two-fold degenerate for each $k$).

At this stage, we have not yet identified
the symmetry class to which the Dirac Hamiltonian
(\ref{eq: 3D Dirac}) belongs;
the interpretation of
the two gradings represented by
a pair of standard Pauli matrices,
$\sigma_{\mu}$ and $\tau_{\mu}$
needs to be specified.
As we will see, the four-component Dirac Hamiltonian
(\ref{eq: 3D Dirac})
realizes topological insulators in classes AII,
AIII, and DIII.

\textbf{Class AII (symplectic)}\,
As discussed by Bernevig and Chen,
\cite{commentBernevigAndChen}
the 3D Dirac Hamiltonian
(\ref{eq: 3D Dirac in k})
is a topological insulator in class AII
since it satisfies
$
{i}\sigma_y \mathcal{H}^*(k) (-{i}\sigma_y)
=
\mathcal{H}(-k)
$,
which we can interpret as a TRS for
half-integer spin.

\textbf{Class DIII}\,
In addition to TRS,
the Dirac Hamiltonian also satisfies PHS,
$
\tau_y\otimes \sigma_y
\mathcal{H}^*(k)
\tau_y\otimes \sigma_y
=
-\mathcal{H}(-k)
$.
Since 
$(\tau_y\otimes \sigma_y)^T
=
\tau_y\otimes \sigma_y
$,
the 3D Dirac insulator (\ref{eq: 3D Dirac}) can
be thought of as a member of class DIII.
It is possible to unitary transform the 
Hamiltonian (\ref{eq: 3D Dirac in k})
by 
$\mathcal{H} \to 
\mathrm{diag}\,(\sigma_0, -{i}\sigma_y)\,
\mathcal{H}
\mathrm{diag}\,(\sigma_0, +{i}\sigma_y)\,
$,
yielding
\begin{eqnarray}
\mathcal{H}(k)
=
\left(
\begin{array}{cc}
m & k\cdot \sigma ({i}\sigma_y)\\
(-{i}\sigma_y)k\cdot \sigma & -m
\end{array}
\right),
\label{eq: class DIII 3D Dirac}
\end{eqnarray}
such that the PHS takes
on
the canonical 
form displayed
in Eq.\ (\ref{eq: def PHS (triplet)}).
One can easily check that
$
{i}\sigma_y \mathcal{H}^*(k) (-{i}\sigma_y)
=
\mathcal{H}(-k)$
and
$
\tau_x \mathcal{H}(k) \tau_x
=
-\mathcal{H}^*(-k).
$
This topological insulator,
Eq.\ (\ref{eq: class DIII 3D Dirac}),
describes
the fermionic BdG quasiparticles in 
the BW state realized 
in the B phase of liquid ${ }^{3}\mathrm{He}$,
for which the
$d$-vector is parallel to the momentum,
$\boldsymbol{d}_k \propto k$.
The $k$-dependence of
the single particle dispersion $\varepsilon_k$ 
of ${ }^{3}\mathrm{He}$
[see Eg.\ (\ref{eq: BdG triplet in k space})]
is weaker around $k=0$
as compared to that of the $d$-vector
and hence neglected.
I.e., the Dirac mass $m$ here is given by
the minus of the chemical potential $\varepsilon_F$,  
$m=\varepsilon_{k=0}=-\varepsilon_F$.
\cite{VolovikBooks}

\textbf{Class AIII}\,
The 3D Dirac Hamiltonian (\ref{eq: 3D Dirac})
can be viewed as an insulator in class AIII
due to the
(chiral)
 symmetry\cite{Zirnbauer96}
$\tau_y \mathcal{H}\tau_y = -\mathcal{H}$.
We can bring the 3D Dirac Hamiltonian (\ref{eq: 3D Dirac})
into block off-diagonal form
by a rotation $\tau_y\to \tau_z$,
which transforms the Dirac mass term in Eq.\ (\ref{eq: 3D Dirac})
into the chiral mass term,
\begin{eqnarray}
\mathcal{H} =
-{i}\alpha_{\mu}\partial_{\mu}
-{i}\beta \gamma^5 m,
\quad
\mu=x,y,z.
\label{eq: AIII Dirac}
\end{eqnarray}
In momentum space,
\begin{align}
\mathcal{H}(k)
&=
\left(
\begin{array}{cc}
0 & k\cdot \sigma-{i}m\\
k\cdot \sigma +{i}m& 0
\end{array}
\right).
\label{eq: class AIII Dirac}
\end{align}
The chiral symmetry is imposed by
$\beta \mathcal{H}(k) \beta = -\mathcal{H}(k).$

\subsubsection{Wave functions, projector and winding number}

It is well known that
the 2D two-component massive Dirac Hamiltonian,
$
\mathcal{H}^{\mbox{\begin{tiny}2D\end{tiny}}}_{\mbox{\begin{tiny}Dirac\end{tiny}}}(k_x,k_y)=
k_x \sigma_x + k_y \sigma_y + m\sigma_z
$,
is the simplest 
example of a topological insulator in 2D
\cite{Deser82,Haldane88,Ludwig94}:
It is an IQH insulator
characterized by the non-trivial Chern integer ($\sigma_{xy}$),
$\sigma_{xy}= \mathrm{sgn}(m)/2 \times (e^2/h)$.
If $(k_x,k_y,m)$ is viewed as a set of parameters
that can be changed adiabatically,
the 2D massive Dirac Hamiltonian 
is nothing but
the $2\times 2$ Hamiltonian considered by Berry himself
to illustrate the Abelian geometric (Berry) phase.
\cite{Berry84}
As described below, 
the $4\times 4$ Dirac Hamiltonian (\ref{eq: 3D Dirac in k})
can be thought of as a natural generalization
of the $2\times 2$ example 
$\mathcal{H}^{\mbox{\begin{tiny}2D\end{tiny}}}_{\mbox{\begin{tiny}Dirac\end{tiny}}}(k_x,k_y)$, 
and exhibits a non-trivial
 non-Abelian Berry phase.
\cite{Chruscinski,Biswas89, Arodz89, Hatsugai04, Chyh-Hong04}

In particular, the two eigenfunctions of the Hamiltonian in
Eq.\ (\ref{eq: 3D Dirac in k})
at wavevector $k$
 with negative energy $E(k)=-\lambda(k)$ are
given by
\begin{align}
|u_1(k)\rangle
&=
\frac{1}{
\sqrt{
2\lambda (\lambda + m)
}
}
\left(
\begin{array}{c}
-k_- \\
k_z \\
0 \\
\lambda+m
\end{array}
\right),
\nonumber \\%%%%%
|u_2(k)\rangle
&=
\frac{1}{
\sqrt{
2\lambda (\lambda + m)
}
}
\left(
\begin{array}{c}
-k_z \\
-k_+  \\
\lambda+m \\
0 \\
\end{array}
\right),
\label{eq: 4-component Dirac wfn negative}
\end{align}
whereas the eigenfunctions with
positive energy $E(k)=+\lambda(k)$ are
\begin{align}
|u_3(k)\rangle
&=
\frac{1}{
\sqrt{
2\lambda (\lambda - m)
}
}
\left(
\begin{array}{c}
k_- \\
-k_z \\
0 \\
\lambda-m
\end{array}
\right),
\nonumber \\%%%%%
|u_4(k)\rangle
&=
\frac{1}{
\sqrt{
2\lambda (\lambda - m)
}
}
\left(
\begin{array}{c}
k_z \\
k_+ \\
\lambda-m \\
0
\end{array}
\right),
\label{eq: 4-component Dirac wfn positive}
\end{align}
where $k_{\pm}=k_x\pm {i}k_y$.
Note that if $m>0$, $|u_{3,4}(k)\rangle$
are not well-defined
at $\lambda(k) = m$ (i.e, at $k=0$).

Correspondingly, the projector ($Q$-matrix) onto
the lowest two negative energy states is given by
\begin{eqnarray}
Q(k) = 2P(k) -1
=
\frac{-1}{\lambda}\left(
k_{\mu}\alpha_{\mu} + m\beta
\right).
\end{eqnarray}
By using the chiral grading
[see Eq.\ (\ref{eq: AIII Dirac})],
we can
define the $q$-matrix 
for class AIII Dirac Hamiltonian
(\ref{eq: class AIII Dirac}) as
\begin{eqnarray}
q(k) =
\frac{-1}{\lambda}\left(
k_{\mu} \sigma_{\mu} -{i}m
\right).
\label{eq: q-matrix for 3D 4 component Dirac}
\end{eqnarray}
[Similarly, for the class DIII massive Dirac Hamiltonian
(\ref{eq: class DIII 3D Dirac}),
the projector is given by
$q(k) =
{i}\sigma_y
\left(
k_{\mu} \sigma_{\mu} -{i}m
\right)/\lambda$,
which satisfies 
$q^T(-k) = q(k)$, 
in the basis that makes the Hamiltonian block off-diagonal
as discussed in 
Eq.\ (\ref{eq: chiral basis for DIII}).]
The winding number $\nu$ 
[Eq.\ (\ref{eq: def winding number})]
for the map represented by $q(k)$
from $S^3$ to $\mathrm{U}(2)$ can be computed as
\begin{eqnarray}
\nu[q] =
\frac{1}{2}
\frac{m}{|m|}.
\label{eq: winding number for 3D top. AIII and DIII}
\end{eqnarray}
The appearance of 
a half-integer value for $\nu$
is common to the continuum descriptions,
and must be supplemented by information about
the structure of wave functions
at high energy (located away from the Dirac point in the BZ).
See, e.g., the discussion of this issue by Haldane 
in the context of the IQHE. \cite{Haldane88}]

For the lower two occupied bands,
we can introduce a U(2) gauge field by
\cite{Wilczek and Zee}
\begin{eqnarray}
A^{\hat a \hat b}_{\mu}(k)dk_{\mu} 
\!\!&=&\!\!
\langle u^{\ }_{\hat a}(k) |d u^{\ }_{\hat b}(k) \rangle,
\quad \hat a, \hat b=1,2,
\label{eq: su2 gauge}
\end{eqnarray}
which can be decomposed into U(1) ($a^0$)
and SU(2) ($a^{j=x,y,z}$)
parts as
\begin{eqnarray}
A^{\ }_{\mu}(k)
=
a^0_{\mu}(k) \frac{\sigma_0}{2{i}}
+
a^j_{\mu}(k) \frac{\sigma_j}{2{i}}.
\label{eq: decomposition of U(2) into U(1)xSU(2)}
\end{eqnarray}
While the U(1) part is trivial,
the SU(2) part is given by
\begin{eqnarray}
a^{i}_{j}(k)
=
-
\epsilon_{i j l}
\frac{k_{l}}{\lambda
(\lambda+m)},
\end{eqnarray}
where 
$i=x,y,z$ and $j,l =x,y,z$.
We have flipped the sign of $k_x$, $k_x \to -k_x$
for notational convenience.

\subsubsection{Boundary Dirac fermions}

We have mentioned above that,
quite generally,
the non-trivial topological properties of the
bulk wave function
manifest  themselves as a gapless
surface state when we terminate the
3D bulk sample by a 2D boundary.
To see this explicitly, 
let us take the mass term to be $z$-dependent
($m>0$),
\begin{equation}
m(z) \to
\left\{
\begin{array}{cc}
+m, & z\to +\infty,\\
\\
-m, & z\to -\infty,
\end{array}
\right.
\label{eq: z-dep mass}
\end{equation}
and look for 2D Dirac fermion solutions
localized at the boundary $z=0$.%
\cite{Callan85,Fradkin86,Boyanovsky87,Kaplan92,
SalomaaVolovik1988,GrinevichandVolovik1988}
For convenience,
 we take the following representation
of the massive 3D Dirac Hamiltonian in class AIII,
$\mathcal{H} =
-{i}\alpha_{\mu}\partial_{\mu}
-{i}\beta \gamma^5 m(z)
$.
The solution to the 3D Dirac equation with energy $E(k_{\perp})$ is
\begin{equation}
\Psi(z)
=
\left(
\begin{array}{c}
0 \\ a(k_\perp)  \\ b(k_\perp) \\ 0
\end{array}
\right)
e^{-\int^z dz' m(z')},
\end{equation}
where $k_{\perp}=(k_x,k_y)$ 
and 
$x_{\perp}=(x,y)$
represent
the momentum and coordinates along the surface,
respectively,
and $a(k_{\perp})$ and $b(k_{\perp})$ are obtained from
the solution to the 2D Dirac equation,
\begin{eqnarray}
\left(
\begin{array}{c}
a(k_{\perp}) \\
b(k_{\perp})
\end{array}
\right)
=
\frac{e^{{i}k_{\perp}\cdot x_{\perp}}}
{\sqrt{2}}
\left(
\begin{array}{c}
e^{{i}\mathrm{arg}\, k_+} \\
\pm  1
\end{array}
\right),
\end{eqnarray}
with
$E(k_{\perp})=\pm \sqrt{k_{x}^2+k_{y}^2}$, respectively.

Below, we will study the stability of
this gapless boundary 2D Dirac state against
the opening of a gap,
by perturbing the Hamiltonian
by static and  homogeneous potentials 
which respect the discrete symmetries
defining the respective symmetry classes.

\textbf{Classes AII and DIII}\,
The gapless nature of the single surface Dirac fermion is
protected by TRS since the opening of a gap would
violate Kramers theorem.
Indeed, for class AII, the only
spatially homogeneous perturbation compatible with the TRS
is a constant scalar potential $V$,
\begin{eqnarray}
\mathcal{H}=
-{i}\partial_{\mu}\sigma_{\mu} + V,
\quad
\mu=x,y,
\label{eq: surface Dirac AII}
\end{eqnarray}
which is known not to open a gap.
For class DIII, on the other hand,
even the scalar potential (chemical potential) $V$ is prohibited
because of PHS.

The stability of the gapless nature of
the single surface Dirac fermion
is guaranteed by the bulk $\mathbb{Z}_2$ invariant
in the symplectic symmetry class (class AII).
\cite{Roy3d, Moore06,Fu06_3Da,Fu06_3Db}
Although this protection of the gapless spectrum by the
$\mathbb{Z}_2$ invariant
also extends to a surface Dirac fermion 
(which is actually Majorana because of PHS in the BdG equation)
in class DIII
when the number of surface Dirac (Majorana)
fermions is odd,
it is only the non-trivial winding number $\nu$
in class DIII
that guarantees
the stability of
an arbitrary number of gapless surface Dirac (Majorana)
fermions
against perturbations (uniform and random).

\textbf{Class AIII}\,
A single flavor
of 2D Dirac fermions
 in class AIII can be perturbed by a
static and homogeneous vector potential:
\begin{eqnarray}
\mathcal{H}&=
-{i}\partial_{\mu}\sigma_{\mu} + A_{\mu}\sigma_{\mu},
\quad
\mu=x,y.
\label{eq: single Dirac + gauge potential}
\end{eqnarray}
The vector potential perturbation
shifts the location of the node, but does not
open a gap.

Although there is a ``hidden'' TRS in class AIII,
the stability of this single Dirac fermion
(\ref{eq: single Dirac + gauge potential})
is not protected by the $\mathbb{Z}_2$ invariant.
This is so since in order to reveal the TRS,
we need to consider the full BdG Hamiltonian
$\mathcal{H}_4$ in (\ref{eq: BdG hamiltonian}),
rather than $\mathcal{H}_2$
defined in (\ref{H_2}) and (\ref{TRS for AIII}).
In $\mathcal{H}_4$, the number of flavors
of the surface Dirac fermions is counted as two,
and not protected by
the $\mathbb{Z}_2$ invariant.
Again, it is the winding number
$\nu$
that guarantees
the stability of
an arbitrary number of flavors of gapless surface Dirac fermions
against perturbations.

\subsection{3D 8-component Dirac Hamiltonian}
\label{sec: 3D 8-component Dirac Hamiltonian}

It turns out that in general we cannot have
a (3+1)D 4-component Dirac Hamiltonian
which is a member of class CI/CII,
and which also 
possesses a gapless Dirac fermion surface mode.
We are thus led to consider
a (3+1)D 8-component Dirac Hamiltonian.
(It is possible to construct
gapless four-component Dirac Hamiltonian
in classes CI and CII, but we cannot give a mass
to them.)

\textbf{Class CI}\,
The following massive 3D 8-component Dirac Hamiltonian
\begin{equation}
\mathcal{H} =
\left(
\begin{array}{cc}
0 & D \\
D^{\dag} & 0
\end{array}
\right),
\quad
D =
{i}\sigma_y
\beta 
\left(
 k_{\mu}\alpha_{\mu} - {i}m \gamma^5
\right),
\label{eq: 8x8 CI 3D Dirac}
\end{equation}
is a member of class CI
since $D^T(k) = D(-k)$.
[See Eq.\ (\ref{eq: chiral rep for class CI}).]
The energy spectrum at
wavevector $k$ is given by
$
E(k)=\pm \sqrt{k^2+m^2}
=\pm \lambda(k)
$,
where each eigenvalue is four-fold degenerate.

The projector takes 
on block off-diagonal form  and is given by
\begin{align}
Q(k) &= 2P(k) -1
=
-\frac{1}{\lambda} \mathcal{H}(k),
\nonumber \\%%%%%
q(k) &=
-\frac{1}{\lambda}
{i}\sigma_y \beta
\left(
 k_{\mu} \alpha_{\mu} -{i}m \gamma^5
\right).
\end{align}
The winding number can be computed as
\begin{eqnarray}
\nu[q] = \frac{1}{2}\frac{m}{|m|}
\times 2,
\end{eqnarray}
which is twice as large as
the winding number for the four-component case.
As before, this winding number should be interpreted
either
$
m/|m|
\times 2$
or 0, depending on the behavior of the wave function
at higher energy.

When we terminate the 3D Dirac insulator
(\ref{eq: 8x8 CI 3D Dirac}),
by a 2D boundary, by making the mass term $z$-dependent
as in Eq.\ (\ref{eq: z-dep mass}),
we find two flavors of surface Dirac fermions,
\begin{align}
\mathcal{H} &=
\left(
\begin{array}{cc}
0 & D^{\ } \\
D^{\dag} & 0
\end{array}
\right),
\nonumber \\%%%%%
D &=
{i}\sigma_y
\left(
k_+
+
A_x \sigma_x
+
A_y \sigma_y
+
A_z \sigma_z
\right),
\label{eq: surface Dirac CI}
\end{align}
where we have included
perturbations $A_{x,y,z}\in \mathbb{C}$
allowed by class CI symmetries.
One can easily check that
$D^{T}(k) = D(-k)$.
The gapless nature of
this four-component Dirac fermion is
stable against arbitrary values of
3 complex (6 real) parameters $A_{x,y,z}$.
Indeed, just like a vector potential perturbation
in class AIII,
$A_{x,y,z}$ shifts the location of the Dirac node
from $(0,0)$ to $(k^{0}_{x},k^{0}_y)$,
where $(k^{0}_{x},k^{0}_y)$ is a solution to
\begin{align}
(k^0_x)^2 - (k^0_y)^2 - \mathrm{Re}A^2 + \mathrm{Im}A^2 &= 0,
\nonumber \\%%%%%
k^0_x k^0_y + (\mathrm{Re}A\cdot \mathrm{Im}A) &= 0.
\end{align}

\textbf{Class CII}\,
The following 8-component
3D Dirac Hamiltonian,
\begin{eqnarray}
\mathcal{H}
=
\left(
\begin{array}{cc}
0 & D \\
D^{\dag} & 0 \\
\end{array}
\right),
\quad
D =
k_{\mu} \alpha_{\mu}
+
m \beta
=D^{\dag},
\label{eq: 8x8 CII 3D Dirac}
\end{eqnarray}
is a member of class CII
since 
$
{i}\sigma_y D^*(k) (-{i}\sigma_y) = D(-k)
$, 
and has a gapped spectrum,
$E(k)=\pm \sqrt{k^2 + m^2}$.
The projector $Q$ and $q$-matrix are given by
\begin{align}
Q(k) & = 2P(k) -1
=
-\frac{1}{\lambda} \mathcal{H}(k),
\nonumber \\%%%%%
q(k) &=
-\frac{1}{\lambda}
\left(
k_{\mu}\alpha_{\mu} +m \beta
\right).
\end{align}

Observe that,
compared with the 
3D Dirac insulator  in class CI,
Eq.\  (\ref{eq: 8x8 CI 3D Dirac}),
the mass term for the 
3D Dirac insulator in 
class CII, Eq.\ (\ref{eq: 8x8 CII 3D Dirac}),
is given by the Dirac mass term ($m\beta$),
not by the chiral mass term (${i}m\gamma^5$).
Due to this difference,
the winding number
vanished for  the
3D Dirac Hamiltonian
in class CII, Eq.\ (\ref{eq: 8x8 CII 3D Dirac}):
\begin{eqnarray}
\nu[q]=0.
\end{eqnarray}

In spite of the vanishing of the
winding number,
we {\it do find} two flavors of
two-component Dirac fermions
at the surface of a 3D Dirac
insulator in 
class CII (\ref{eq: 8x8 CII 3D Dirac}),
when 
making the mass term $z$-dependent
as in Eq.\ (\ref{eq: z-dep mass}).
In particular, consider the
general form of the Dirac Hamiltonian
on the 2D surface \cite{Bernard01,Guruswamy00}
\begin{align}
\label{eq: cII 2D Dirac}
\mathcal{H}
&=
\left(
\begin{array}{cc}
0 & D \\
D^{\dag} & 0
\end{array}
\right),
\\%%%%%
D&=
\left(
\begin{array}{cc}
v_+ & k_-+a \\
k_+-\bar{a} & v^*_+
\end{array}
\right)
\nonumber\\
&=
\left(k_x + {i}a_x\right)\sigma_x
+
\left(k_y + {i}a_y\right)\sigma_y
\nonumber \\%%%%%
&
\quad
+
\mathrm{Re}\,v_+ \sigma_0
+
{i}\,\mathrm{Im}\,v_+ \sigma_z,
\nonumber
\end{align}
where the perturbations
$a_{x,y}$ and $v_+$
are the only ones allowed by class CII symmetries.
(We used the notation $a=\mathrm{Re}\,a+{i}\mathrm{Im}\,a=a_y+{i}a_x$.)
It turns out that the gapless nature of
these surface Dirac fermions
is preserved by these perturbations.
To see this, consider the determinant of the Hamiltonian
\begin{eqnarray}
\mathrm{det}\left(DD^{\dag}\right)
=
\left(
|v_+|^2-|k_+|^2+|a|^2\right)^2
-
\left(
\bar{a} k_- -a k_+
\right)^2,
\nonumber \\%%%%%
\end{eqnarray}
which vanishes when
\begin{eqnarray}
|k_+ |^2=|v_+|^2+|a|^2
\quad \mbox{and}\quad
(k_x,k_y)\perp (a_x,a_y).
\end{eqnarray}
This shows that it is always possible to find a
wavevector $(k_x,k_y)$ for which the determinant,
and thus the energy eigenvalue vanishes,
proving the absence of a gap.

Therefore we conclude that
the 3D Dirac insulator (\ref{eq: 8x8 CII 3D Dirac})
is a non-trivial topological insulator in class CII:
it is impossible to deform
the insulator (\ref{eq: 8x8 CII 3D Dirac})
into a topologically trivial insulator
(an insulator without a stable surface state)
without closing the energy gap in the 3D bulk,
because the existence of the gapless 
Dirac fermion surface modes
plays the role of a topological invariant.
On the other hand, when the number of
flavors is twice an even integer,
one can easily find a perturbation that gives
a mass gap to all surface Dirac fermions.

%%%%%%%%%%%%%%%%%%%%%%%%%%%%%%%%%%%%%%%%%%%%%%%%%%%%%%%%%%%%%%%%%%%
\section{Topological field theory description}
\label{sec: topological field theory description}

%Feynman's slash
\newcommand{\Slash}[1]{\ooalign{\hfil/\hfil\crcr$#1$}}

In order to understand more  intuitively
the reason why symmetry classes with a sublattice (chiral) symmetry
(classes AIII, DIII, and CI) possess
stable gapless surface Dirac fermion modes,
we derive in this section
a \textit{doubled}-Chern-Simons field theory
describing the 3D bulk insulator.
To this end
we identify, 
following
the spirit of Read and Green,\cite{Read00}
conserved charges of
the action,
and introduce \textit{external} gauge fields
that couple minimally to these charges.
The gapped fermionic degrees of freedom in the
3D bulk are then integrated out
to derive the effective action of the gauge fields.

A similar procedure has also been discussed
for domain-wall fermions~\cite{Kaplan92} in 
lattice gauge theory, where
a (\textit{non-doubled}) Chern-Simons theory can be derived for
$[(2n-1)+1]$D boundary fermions
of the $(2n+1)$D bulk.\cite{Golterman92}

As an example, let us take the class AIII Dirac insulator
(\ref{eq: AIII Dirac}) in three spatial dimensions.
The generating function for
the single-particle Green's function
can be written as
a fermionic functional integral,
\begin{align}
Z &= \int\mathcal{D}\left[\psi^{\dag},\psi\right]e^{-S},
\nonumber \\%%%%%
S
&=
{i}
\int d^3x\,
\psi^{\dag}
\left(
\mathcal{H} -
{i}\eta
\right)
\psi.
\label{eq: generating funct. for Green functions}
\end{align}
Here, note that we are using a
three-dimensional (Euclidean) action,
instead of a (3+1)-dimensional one (compare, e.g., with
Ref.\ \onlinecite{Xiao-LiangQi08}),
since we are focusing on
single-particle Green's functions in the absence of interactions.
A finite 
level-broadening term $\eta \neq 0$ is necessary 
to regularize
delta functions 
appearing
in 
the single particle Green's function.

The action enjoys a (electromagnetic) U(1) symmetry
\begin{eqnarray}
\psi^{\dag} \to \psi^{\dag}
e^{+{i} \theta},
\quad
\psi \to e^{-{i} \theta} \psi,
\label{eq: U(1) symmetry}
\end{eqnarray}
where $\theta\in \mathbb{R}$.
Due to the sublattice (or chiral) symmetry,
$\beta \mathcal{H}\beta = -\mathcal{H}$,
the functional integral 
possesses the additional symmetry
\begin{eqnarray}
\psi^{\dag} \to \psi^{\dag}
e^{+{i}\beta \theta},
\quad
\psi \to e^{+{i}\beta \theta} \psi, 
\label{eq: axial U(1) symmetry}
\end{eqnarray}
when $\eta=0$.
A finite level-broadening term
$\eta$ spoils this symmetry.
Instead of adding
a level-broadening term, however,
regularization of the functional
integral can also be achieved,
alternatively, by attaching ideal leads
(or perfect absorbers) respecting the chiral symmetry
to the sample.

We now proceed to derive the Chern-Simons theory.
Corresponding to the two continuous symmetries
discussed above,
we couple two U(1) gauge fields $a_{\mu}$ and $b_{\mu}$
to the
fermions,
\begin{equation}
S
=
\int d^3x\,
\bar{\psi}
\left(
\Slash{\partial}
-
{i}
\Slash{a}
-
{i}
\gamma^0
\Slash{b}
+
m \gamma_5
\right)
\psi,
\end{equation}
where we have introduced 
the abbreviations
$\bar{\psi}=\psi^{\dag}\beta$,
$\gamma^0=\beta$, 
$\gamma^k = \beta \alpha_k$,
and
$\Slash{a}:= \gamma^{\mu}a_{\mu}$.
While the external U(1) gauge field $a_{\mu}$,
associated with the global U(1) symmetry of Eq.\ (\ref{eq: U(1) symmetry}),
couples to the electromagnetic current,
the external ``axial'' gauge field $b_{\mu}$ detects 
the \textit{sublattice-resolved} current,
as it is associated with the global U(1) symmetry transformation defined
in Eq.\ (\ref{eq: axial U(1) symmetry}),
where equal and opposite U(1)
transformations are performed 
on the two sublattices $A$ and $B$ of the
underlying bipartite lattice.

To be more general, we discuss the case of $N$ replicas
of the above 3D Dirac fermions,
and couple them minimally with two $\mathrm{U}(N)$ gauge fields,
$a_{\mu} = a^a_{\mu} T_a$ and $b_{\mu}= b^a_{\mu} T_a$, 
with the generators $T_a$. 
The use of replicas is a convenient method to compute
disorder averaged physical quantities in the presence of random impurities.
Since we only intend to give a schematic derivation of 
the doubled Chern-Simons theory, we do not add any explicit disorder
potential.

We now integrate out the fermions and derive the effective action
for the gauge fields $a_{\mu}$ and $b_{\mu}$,
\begin{eqnarray}
\int 
\mathcal{D}\left[\bar{\psi},\psi\right]
e^{-S}
=
e^{-S_{\mathrm{eff}}\left[ a_{\mu},b_{\mu} \right]},
\end{eqnarray}
by a derivative expansion
\begin{align}
S_{\mathrm{eff}}
&=
-\mathrm{Tr}\,
\ln
\left( G^{-1}_0 -V \right)
\nonumber \\%%%%%
&=
-\mathrm{Tr}\,
\ln G^{-1}_{0}
+
\sum_{n=1}^{\infty}
\frac{1}{n}
\mathrm{Tr}\,
\left(G_0 V\right)^n,
\end{align}
where $G_0$ denotes
the propagator of free 3D Dirac fermions,
which is given
in momentum space by
\begin{align}
G_0(k) &=
-\frac{{i}\Slash{k}+ m \gamma_5 }{k^2+m^2},
\end{align}
whilst
\begin{align}
V(q) &=
-{i}\Slash{a}_{\mu}(q)
-{i}\gamma^0 \Slash{b}_{\mu}(q).
\end{align}

Introducing
the linear combinations
\begin{eqnarray}
A^{\pm}_{\mu}= a_{\mu} \pm b_{\mu},
\end{eqnarray}
the resultant effective action, to leading order
in the derivative expansion,
takes the form of
a (Euclidean) doubled Chern-Simons theory,
\begin{align}
S_{\mathrm{eff}}
&=
\frac{1}{2}
\frac{m}{|m|}
\left(
I[A^+]
-
I[A^-]
\right)
+
\mbox{div.},
 \\%%%%%
I[A]
&=
\frac{-{i}}{4 \pi}
\int d^3x\,
\epsilon^{\mu\nu\lambda }
\mathrm{tr}
\left(
A^{\ }_{\mu}
\partial^{\ }_{\nu} A^{\ }_{\lambda}
+
\frac{2{i}}{3}
A^{\ }_{\mu}A^{\ }_{\nu}A^{\ }_{\lambda}
\right),
\nonumber
\end{align}
where ``$\mbox{div.}$'' represents
an ultraviolet (UV)
linearly divergent piece.
This divergence is closely related to the
appearance of the half-integer
coefficient of the Chern-Simons term,
$\mathrm{sgn}(m)\times 1/2$:
the action
$\pm \mathrm{sgn}(m)I[A]/2$ is not gauge invariant by itself.
[See also
the
discussion below Eq.\ (\ref{eq: winding number for 3D top. AIII and
DIII}).]

Introducing a gauge invariant regulator,
such as the 
Pauli-Villars (PV)
regularization,
cures both 
the UV divergence
and the half-integer coefficient.
\cite{Redlich84,Dunne99}
Here, note that, since there is no chiral anomaly in 3D,
the functional integral can be regularized
without breaking the two $\mathrm{U}(N)$ gauge symmetries,
although the parity symmetry can be destroyed by the
regularization.
In the Pauli-Villars regularization, we define the physical,
divergence-free effective action $S^{\mathrm{PV}}_{\mathrm{eff}}$ by
\begin{eqnarray}
S^{\mathrm{PV}}_{\mathrm{eff}} = S_{\mathrm{eff}}(0) -
\lim_{M^2\to \infty} S_{\mathrm{eff}}(M),
\end{eqnarray}
where $S_{\mathrm{eff}}(M)$ represents the effective action
in the presence of two massive Dirac particles:
 the original
particle with the mass $m$ and 
another one
with mass $M$, which we take
to infinity ($M^2\to \infty$).
Here, the second particle (which is bosonic)
might be interpreted as supplementing
the missing information
far away from the Dirac point discussed
around Eq.\ (\ref{eq: winding number for 3D top. AIII and DIII}).
The coefficient of the Chern-Simons terms 
$I[A^+]$ and $I[A^-]$
in $S^{\mathrm{PV}}_{\mathrm{eff}}$,
which is $(1/2)(m/|m|-M/|M|)$ instead of $(1/2) m/|m|$,
depends on the sign of the regulator mass $M$: 
when $\mathrm{sgn}(M)=-\mathrm{sgn}(m)$,
this coefficient  equals
$\mathrm{sgn}(m)$ whereas it
vanishes
when $\mathrm{sgn}(M)=+\mathrm{sgn}(m)$.
These two cases represent the topological non-trivial and trivial phases,
respectively.

Once we have  established  the appearance of the 
doubled Chern-Simons term for the resulting 
3D bulk theory, we conclude that
the surface degrees of freedom,
which appear when the 3D bulk is terminated by
a 2D surface,
are described by
the 
two-dimensional
$\mathrm{U}(N)$ Wess-Zumino-Witten (WZW) theory
at level $k=1$.
The WZW theory is  well known to 
be gapless and to possess
both, holomorphic and antiholomorphic 
sectors.\cite{KnizhnikZamolodchikov}
Thus, the $\mathrm{U}(N)\times \mathrm{U}(N)$ symmetry of
the gapped 3D bulk theory,
which represents two independent transformations
for each sublattice, 
turns into the two independent
holomorphic and
antiholomorphic $\mathrm{U}(N)$ gauge symmetries
of the WZW theory describing the resulting
degrees of freedom  
at the surface.

An entirely analogous discussion
can be carried through for the 3D
topological insulators in symmetry classes 
DIII and CI,
for which the relevant gauge groups
appearing in the gapless WZW theory at the 2D
surface are
$\mathrm{O}(2N)$ and $\mathrm{Sp}(N)$, respectively.

%%%%%%%%%%%%%%%%%%%%%%%%%%%%%%%%%%%%%%%%%%%%%%%%%%%%%%%%%%%%%%%%%%%

\section{Many-body aspects of 3D topological superconductors}

\label{sec: 3D superconductors as a topological phase}

Up to now, we have treated 
the three-dimensional superconductors
at the mean-field level of pairing
(where the gap function $\Delta $ is
``frozen'' to be a constant, i.e., frequency independent):
we have focused solely on the
dynamics of the fermionic BdG quasiparticles,
existing deep within the superconducting state.
In the full description of superconductivity,
however, the 
gap function $\Delta$,
together with the electromagnetic U(1) gauge field, 
have to be regarded as  
dynamical entities.
Once we employ this full description,
the superconductors of interest in this paper
should be regarded as topological phases with
non-trivial ground state degeneracy
(and charge fractionalization).
\cite{Hansson04,Oshikawa06}
(Specifically,
while the BdG quasiparticle sector 
of the B-phase ${ }^{3}\mathrm{He}$
is a topological Dirac insulator in class DIII 
[as discussed in Eq.\ (\ref{eq: class DIII 3D Dirac})], 
the B-phase of liquid ${ }^{3}\mathrm{He}$
is in reality not a 3D topological phase, as it supports
gapless Nambu-Goldstone bosons.)

\subsection{Many-body wave function in real space}
Let us first take a closer look at the
many-body ground state wave function of the
three-dimensional class DIII 
Dirac insulator
[see e.g.,  Eq.\ (\ref{eq: class DIII 3D Dirac})].
In a triplet superconductor,
the BCS ground state
$|\mathrm{BCS}\rangle$
projected onto
a space of fixed 
electron number $N$,
$|x_1,\sigma_1;x_2,\sigma_2;\ldots; x_N,\sigma_N
\rangle$
(where $\sigma_i$ represents spin coordinate),
is given by
the wave function
\begin{eqnarray}
&&
 \Psi(x_1,\sigma_1;x_2,\sigma_2;\ldots; x_N,\sigma_N)
\nonumber \\%%%%%
&&  \qquad
:=
\langle x_1,\sigma_1;x_2,\sigma_2;\ldots; x_N,\sigma_N |
\mathrm{BCS}\rangle
\nonumber \\%%%%%
&& \qquad
=
\mathrm{Pf}\,
\left[
g(x_i,\sigma_i;x_j,\sigma_j)
\right],
\end{eqnarray}
where $\mathrm{Pf}$ denotes the Pfaffian 
of the matrix 
$g_{i,j}:=g(x_i,\sigma_i;x_j,\sigma_j)$.\cite{Read00}
The Fourier transform of
$g(x,\sigma,y,\tau)=
g_{\sigma\tau}(x-y)$
as obtained from
(\ref{eq: 4-component Dirac wfn negative})
and
(\ref{eq: 4-component Dirac wfn positive}),
reads
\begin{eqnarray}
\label{gOfk}
g(k) =
(-\lambda+m)
\frac{(k\cdot \sigma){i}\sigma_y }{2k^2}.
\end{eqnarray}
Noting that
$
\lambda(k)
=
\sqrt{k^2+m^2}
\to
|m| + |k|^2/(2|m|)
+\cdots
$
in the long-wavelength limit,  $k\to 0$,
the expression 
in Eq.\ (\ref{gOfk})
takes in that limit the form
\begin{eqnarray}
g(k) \sim
\left\{
\begin{array}{ll}
\displaystyle
- \frac{(k\cdot \sigma){i}\sigma_y}{4|m|}, & m>0, \\
& \\
\displaystyle
-|m| \frac{(k\cdot \sigma){i}\sigma_y}{k^2}, & m<0.
\end{array}
\right.
\end{eqnarray}
Correspondingly, the real-space wave function
$g(r)
=
(2\pi)^{-3}
\int d^3 k\,
e^{{i}k\cdot r} g(k)
$
takes {\it at long scales}
the following form
\begin{eqnarray}
g(r)
\sim
\left\{
\begin{array}{cc}
\displaystyle
\frac{\sigma_{\mu}{i}\sigma_{y}}{4|m|}
{i}\partial_{\mu}\delta^{(3)}(r),
& m >0,\\
& \\
\displaystyle
-|m|
\frac{({i}\sigma \cdot r)({i}\sigma_y)}{4\pi r^3},
& m <0.
\end{array}
\right.
\end{eqnarray}
This behavior is similar to
the strong and weak pairing phases of
the (2+1)-dimensional chiral $p$-wave
superconductor.\cite{Read00}
In one phase,
 the strong pairing phase ($m>0$),
the wave function $g(r)$ of a pair is short-ranged,
whereas in the other, the weak pairing phase ($m<0$),
$g(r)$
exhibits a power-law behavior
and is given by
the correlation function of the two-component
massless 3D Dirac fermion.
Thus, in the weak pairing phase,
the many-body ground state wave function behaves
at large scales as
\begin{align}
&
\Psi(x_1,\sigma_1;x_2,\sigma_2;\ldots; x_N,\sigma_N)
\nonumber \\%%%%%
&
\qquad 
\sim
\mathrm{Pf}
\left[
\frac{
\left(
 \sigma \cdot (x_i-x_j) {i}\sigma_y 
\right)_{\sigma_i \sigma_j}
}
{|x_i-x_j|^3}
\right].
\end{align}
Observe that this is nothing but 
the multi-point correlation function of 
a (simple) 3D conformal field theory,
namely
the 3D free Majorana fermion quantum field theory
defined by the partition function
\begin{eqnarray}
Z=\int \mathcal{D}[\bar{\psi},\psi]
e^{
-\int d^3 x\,\mathcal{L} },
\quad 
\mathcal{L}=
\bar{\psi} \sigma_{\mu} \partial_{\mu} \psi, 
\end{eqnarray}
where $\psi$ is a two-component Grassmann variable
with Majorana condition 
$\bar{\psi}=\psi^T {i}\sigma_y$,
and
$\mu=x,y,z$.
This is analogous to 
the Moore-Read Pfaffian wave function,
which is given by
the multi-point correlation function of 
the 2D Ising conformal field theory
(free Majorana fermion field theory).

\subsection{Ground state degeneracy}

With both the pairing potential $\Delta$ and
the U(1) gauge field being frozen,
there is a unique ground state both in
the strong and weak pairing phases.
We now include quantum fluctuations
of $\Delta$ and the U(1) gauge field.
One consequence of
the inclusion of these
as dynamical degrees of freedom is the
appearance of a non-trivial ground state degeneracy.
The counting of ground states in each phase
is completely parallel to the case of the Moore-Read Pfaffian state
as we will see below.

To count the ground state degeneracy on the
three-torus $T^3$,
we consider 
periodic or anti-periodic boundary conditions (BCs) 
along the three cycles in the  $x,y,z$ directions.
We denote sectors with different BCs 
by $(\iota_x,\iota_y,\iota_z)$,
where $\iota_{\mu}=\pm$
represents periodic/anti-periodic BC.
Following the argument by Read and Green, \cite{Read00}
we notice that
the state with $k=0$ is allowed
only for the $(+,+,+)$ sector,
and it is occupied in the weak pairing
phase whereas it is unoccupied in the strong pairing
phase.

In the strong pairing phase,
different boundary conditions
lead to $2^3$ degenerate ground states,
and all of them have an
even number of fermions.
On the other hand, in the weak pairing phase,
the ground state
 for
the boundary condition sector
 $(+,+,+)$ has
an odd number of particles
because of an additional occupied state at $k=0$.
Thus, the
ground state degeneracy for
an even number of fermions 
is $2^3-1=7$ whereas there is a unique ground state
for 
an odd number of fermions.
This should be contrasted with
the ground state degeneracy of $2^3$ present 
in the 3D Abelian Higgs model
which can be described by
a $(3+1)$D $BF$ topological field theory.
\cite{Hansson04}
It is unclear what kind of bulk topological field
theory can describe the weak pairing phase,
as it has fermionic excitations at boundaries,
unlike the bosonic boundary excitations in the strong pairing phase
described by the $BF$ topological field theory. 

The smaller topological degeneracy in the weak pairing phase
can also be understood in terms of
the ``blocking mechanism'' introduced in
Ref.\ \onlinecite{Oshikawa06}.
The smaller topological degeneracy in the Moore-Read Pfaffian state
happens because a vortex-antivortex excitation carries
a Majorana fermion at the core.
In a topological phase, a different ground state,
starting from a given ground state, can be generated
by first creating a particle-antiparticle pair
out of the ground state,
then moving around the quasi-particle
along a homotopically non-trivial cycle,
and finally pair-annihilating the pairs.
If the quasi-particle accommodates a Majorana fermion,
however,
the last step of the above procedure,
which is pair-annihilation,
 might not be possible
(it might be ``blocked'').
In the 3D Pfaffian state, we have vortex lines, instead of
vortices, which do support Majorana fermion modes.
\cite{VolovikBooks,Misirpashaev95,Callan85,Sato03}
Thus, we expect a similar blocking mechanism
should apply.

We now briefly discuss the effects of interparticle interactions.
Since short-range interactions are irrelevant
by power-counting for free Dirac fermions in $(2+1)$
dimensions, we would expect the gapless fermionic
surface modes in the weak pairing phase
to be stable against
the formation of a gap,
up to, possibly,  some critical interaction strength
(certainly when random disorder potentials are
not simultaneously present).
This should be contrasted with the surface states
in the strong pairing phase,
which are generically gapped as we can see,
for example, from the $BF$ topological field theory.
\cite{Diamantini1996, Hansson04, Diamantini2006}
This should also be compared with the surface states
of three-dimensional
$\mathbb{Z}_2$ topological insulators in the symplectic class (AII),
which are unstable against the BCS pairing instability
because there is a finite Fermi surface (circle),
i.e., finite density of states, within the surface 
Brillouin zone,
for a general value of the chemical potential.

%%%%%%%%%%%%%%%%%%%%%%%%%%%%%%%%%%%%%%%%%%%%%%%%%%%%%%%%%%%%%%%%%%%
\section{Discussion}
\label{sec: discussions}

In this paper, we have undertaken 
the program of classifying 
possible phases of topological insulators
and superconductors
in three spatial dimensions.
Our results have their root in the very general classification
scheme for random matrices obtained 
by Zirnbauer, and Altland and Zirnbauer (AZ) \cite{Zirnbauer96,Altland97}
more than a decade ago,  resulting  in ten  such classes
which extend the well-known three Wigner-Dyson
classes.
Guided by the lessons learned from 
the $\mathbb{Z}_2$ topological insulator
discussed by
Kane and Mele \cite{KaneMele,Fu06_3Da,Fu06_3Db}
and others,\cite{Moore06,Roy3d}
which belongs to the symplectic
(``spin-orbit'')
symmetry class (AII)
in the AZ classification scheme,\cite{Zirnbauer96,Altland97}
we have asked if two quantum ground states in a given symmetry class
can be continuously deformed into each other while keeping 
the  discrete symmetries 
defining the symmetry class intact.
Specifically, we have shown that,
in addition to the three-dimensional 
$\mathbb{Z}_2$ topological insulators
there exist 3D topological insulators
possessing the symmetries of four additional random matrix
classes denoted by AIII, DIII, CI and CII in the work of AZ, 
all of which support 
stable gapless Dirac fermion surface modes
(Majorana fermion surface modes for class DIII).
In particular, we find that the topological properties of the bulk wave
functions in the three symmetry classes AIII, DIII and CI
are characterized by an integral winding number, 
while the bulk characteristics of topological insulators in 
class CII can be described
by a $\mathbb{Z}_2$ number, akin to the well studied topological
insulator in the symplectic symmetry class AII.

\begin{table*}[ht]
\begin{center}
\begin{tabular}{|c||c|c|c|}\hline
AZ class &  Space of transfer matrices & 2D top. insulator & Possible phys. realization 
\\ \hline \hline 
A 
&
$\mathrm{U}(p,q)/\mathrm{U}(p)\times \mathrm{U}(q)$
& 
IQHE ($p\neq q$)
&
GaAs/AlGaAs
\\ \hline
AI 
&
$\mathrm{Sp}(N,\mathbb{R})/\mathrm{U}(N)$ 
& -
& -
\\ \hline
AII (even)
&
$\mathrm{SO}^*(4N)/\mathrm{U}(2N)$
& -
& -
\\ \hline
AII (odd) 
&
$\mathrm{SO}^*(4N+2)/\mathrm{U}(2N+1)$
&
$\mathbb{Z}_2$ top. ins. (QSHE)
&
HgTe/(Hg,Cd)Te
\\ \hline \hline
AIII
& 
$\mathrm{GL}(N, \mathbb{C})/\mathrm{U}(\mathrm{N})$
& -
& -
\\ \hline
BDI
&
$\mathrm{GL}(N,\mathbb{R})/\mathrm{O}(N)$
&-
& -
\\ \hline
CII
&
$\mathrm{GL}(N,\mathbb{H})/\mathrm{Sp}(N)
\equiv
\mathrm{U}^*(2N)/\mathrm{Sp}(N)
$
&-
&-
\\ \hline \hline
D &
$\mathrm{SO}_0(p,q)/\mathrm{SO}(p)\times \mathrm{SO}(q)$ 
&
Thermal QHE ($p\neq q$)
&
Spinless chiral $p$-wave SC
\\ \hline
C 
&
$\mathrm{Sp}(p,q)/\mathrm{Sp}(p)\times \mathrm{Sp}(q)$ 
&
Spin QHE ($p\neq q$)
&
($d \pm i  d$)-wave SC
\\ \hline
DIII (even) &
$\mathrm{SO}(2N,\mathbb{C})/\mathrm{SO}(2N)$ 
&-
&-
\\ \hline
DIII (odd) 
&
$\mathrm{SO}(2N+1,\mathbb{C})/\mathrm{SO}(2N+1)$ 
&
$\mathbb{Z}_2$ top. SC
&
($p+ip$) $\times$ ($p-ip$)-wave SC 
\\ \hline
CI
&
$\mathrm{Sp}(N,\mathbb{C})/\mathrm{Sp}(N)$ 
&-
&-
\\ \hline
\end{tabular}
\caption{
\label{tab: dmpk}
This table lists the space of ensembles of 
(the radial coordinates of) transfer matrices for 
quasi one-dimensional disordered quantum wires
 for each Altland-Zirnbauer (AZ) class.\cite{Caselle06} 
Five of these ensembles of transfer matrices 
describe localization properties of an edge of
a two-dimensional topological insulator or superconductor (SC).
The conventional name of these five 
two-dimensional topological insulators (2D top. ins.)
is given in the third column. The last column
lists some possible physical realizations of these
topological insulators.
}
\end{center}
\end{table*}

\subsection{Topological bulk characteristics and
Anderson delocalization at the boundary}

Another lesson learned from the $\mathbb{Z}_2$ topological insulator 
is an intimate connection between the
topological characteristics 
of the  clean (no disorder, or impurities) system
in the 3D bulk
and the Anderson localization physics
occurring, due to disorder,  at 
two-dimensional boundaries
of such a system:
the
surface
of a
three-dimensional
 $\mathbb{Z}_2$ topological insulator
is a perfect metal ($\mathbb{Z}_2$ topological metal)
in the presence of
disorder which respects the TRS.
This can be understood in terms of the field theoretical framework of 
Anderson localization.
\cite{LeeRamakrishnan, Efetov97}
The fermionic replica non-linear $\sigma$ model (NL$\sigma$M)
describing quantum transport in the
corresponding
symplectic
(``Wigner-Dyson'')
symmetry class
possesses the coset space
$\mathrm{O}(4N)/\mathrm{O}(2N)\times\mathrm{O}(2N)$
as target space.
($N$ is the number of replicas.)
Because 
the homotopy group
$\pi_2[\mathrm{O}(4N)/\mathrm{O}(2N)\times\mathrm{O}(2N)]=\mathbb{Z}_2$,
a $\mathbb{Z}_2$ topological term is allowed in the action
of this NL$\sigma$M,
and
this term is indeed realized at 
a surface of 3D $\mathbb{Z}_2$ topological insulators.
It is the $\mathbb{Z}_2$ topological term
that is responsible for the 
lack of localization and the metallic behavior
at the surface.
\cite{Ando98, Ostrovsky07, Ryu07, Bardarson07, Nomura07}

A key result  of the present paper
is a generalization of these properties of the 3D $\mathbb{Z}_2$ topological  insulator
to 
%the 
topological insulators
belonging to the above mentioned 
four symmetry classes AIII, DIII, CI and CII.
For three of these four classes,
namely for
AIII, DIII, and CI,
which describe the dynamics
of quasiparticles within certain
superconductors,
the corresponding NL$\sigma$M describing Anderson localization
at the surface of the 3D bulk
is the 
principal chiral model (PCM)
on the groups $\mathrm{U}(N)$, 
$\mathrm{O}(2N)$, and $\mathrm{Sp}(N)$,
respectively,
supplemented by a
Wess-Zumino-Witten (WZW) term.
For class CII, the corresponding NL$\sigma$M
is defined on
the coset space
$\mathrm{U}(2N)/\mathrm{O}(2N)$,
which allows
for a $\mathbb{Z}_2$ topological term,
since
the homotopy group
of this space is
$\pi_2[\mathrm{U}(2N)/\mathrm{O}(2N)]=\mathbb{Z}_2$.
Table \ref{tab: nlsm} summarizes the 
target spaces of the corresponding
fermionic replica NL$\sigma$Ms,
as well as
possible 2D topological or WZW terms.
(See, e.g., Refs.\ \onlinecite{Fendley00} and \onlinecite{Fendley01}).
The NL$\sigma$Ms living on the surface of the 3D bulk
``remember'' the non-trivial topological characteristics of the bulk 
through the presence of either a
WZW or a $\mathbb{Z}_2$ topological term.
\cite{commentPruisken}
Due to these WZW or topological terms,
the quantum states existing
at the surface of the 3D topological bulk
are gapless and their gaplessness is
protected against (Anderson) localization
by random potentials respecting the discrete symmetries.
We thus conclude that surfaces 
of 3D topological insulators
with a TRS
are always (``topologically'')  delocalized.
[For sublattice and superconducting classes,
the corresponding
gapless states at the surface
are semi-metal like 
(with conductivities of order unity in natural units;
see end of this section),
unlike
the surface states of $\mathbb{Z}_2$ topological insulators,
which are perfect metals.]

\subsection{Topological insulators in one and two spatial dimensions}

This correspondence between
nontrivial topological  characteristics
of an insulator in the bulk
and lack of
Anderson localization due to random
impurities at the boundary
also applies
to 2D topological insulators and their 1D edge modes
at boundaries.
In (quasi) 1D, Anderson localization problems can be well-described
by the Dorokhov-Mello-Pereyra-Kumar (DMPK) equations,
which is a Fokker-Planck equation describing
the distribution of the eigenvalues of transfer matrices of 
the quasi 1D wire as a function of the wire length.
The ensembles of transfer matrices can be systematically
enumerated, and 
there are twelve possible
DMPK equations.
\cite{Brouwer05,Caselle06}
(See Table \ref{tab: dmpk}.)
The extra two cases which appear here, in addition  to
the ten AZ symmetry classes,
cannot be realized as a quasi 1D tight-binding (lattice) 
model,
but can {\it only} be realized at the
(one-dimensional) 
boundary of 2D topological insulators.
One of these two cases is the symplectic symmetry class (AII),
\cite{Takane04,Zirnbauer92}
which can be realized, for example, at the edge of
the 2D Kane-Mele model.
The other is in class DIII,
\cite{Gruzberg05}
and can be realized, for example, at the edge of
the equal superposition of two chiral $p$-wave
superconductors with 
opposite chiralities [$(p+{i}p)$- and $(p-{i}p)$-wave],
in two spatial dimensions.
\cite{RoySC,SenguptaRoySC2,RoySF2008,Xiao-LiangQiTopSC08}
In classes A (unitary), D, and C,
the DMPK equation depends
on two integers $p$ and $q$, representing the number of
left- and right-moving channels, respectively.
When the numbers of left- and right-moving channels are not equal ($p\neq q$),  
i.e., when the quasi 1D system is chiral,
the corresponding Anderson localization problem 
can only be realized at
an edge of a 2D topological insulator.
Specifically, in class A this topologically 
non-trivial 2D
quantum ground state
is commonly known as the IQHE, 
in class D it is the thermal quantum Hall effect in superconductors,
\cite{Read00,SenthilFisher2000} 
and in class C the spin quantum Hall effect in superconductors,
\cite{SenthlFisherBalentsNayak1998,SenthilMarstonFisher} 
respectively.

\begin{table}[t!]
\begin{center}
\begin{tabular}{|c||c|c|}\hline
AZ class &  Space of Hamiltonians & 1D top.\ ins.
\\ \hline \hline 
A 
&
$\mathrm{U}(N)\times \mathrm{U}(N)/\mathrm{U}(N)
%=\mathrm{U}(N)
$
& 
-
\\ \hline
AI 
&
$\mathrm{U}(N)/\mathrm{O}(N)$ 
& -
\\ \hline
AII 
&
$\mathrm{U}(2N)/\mathrm{Sp}(N)$ 
& -
\\ \hline \hline
AIII
& 
$\mathrm{U}(p+q)/\mathrm{U}(p)\times \mathrm{U}(q)$
& 
$\mathbb{Z}$
\\ \hline
BDI
&
$\mathrm{SO}(p+q)/\mathrm{SO}(p)\times \mathrm{SO}(q)$
&
$\mathbb{Z}$
\\ \hline
CII
&
$\mathrm{Sp}(p+q)/\mathrm{Sp}(p)\times \mathrm{Sp}(q)$
&
$\mathbb{Z}$
\\ \hline \hline
D (even) &
$\mathrm{SO}(N)\times \mathrm{SO}(N)/\mathrm{SO}(N)
%=\mathrm{SO}(N)
$ 
&
-
\\\hline 
D (odd) (``B'') &
$\mathrm{SO}(2N+1)$ 
&
$\mathbb{Z}_2$ 
\\ \hline
C 
&
$\mathrm{Sp}(N)\times \mathrm{Sp}(N)/\mathrm{Sp}(N)
%=\mathrm{Sp}(N)
$ 
&
-
\\ \hline
DIII (even) &
$\mathrm{SO}(2N)/\mathrm{U}(N)$ 
&-
\\ \hline
DIII (odd) 
&
$\mathrm{SO}(4N+2)/\mathrm{U}(2N+1)$ 
&
$\mathbb{Z}_2$ 
\\ \hline
CI
&
$\mathrm{Sp}(N)/\mathrm{U}(N)$ 
&-
\\ \hline
\end{tabular}
\caption{
\label{tab: ivanov}
This table lists the space of ensembles of random matrices for 
zero-dimensional disordered quantum systems
 for each Altland-Zirnbauer (AZ) class.\cite{Caselle06} 
According to Ref.\ \onlinecite{Ivanov99}, 
five of these ensembles of random matrices describe 
localization properties at a (zero-dimensional) boundary of a one-dimensional
topological insulator (1D top.\ ins. ).
}
\end{center}
\end{table}

Finally, a similar correspondence exists also between
one-dimensional topological insulators and their zero-dimensional edges,
where disorder effects can be discussed in terms of 
random matrix theories (RMTs).
Following Ivanov \cite{Ivanov99},
five out of ten AZ classes,  DIII, D, AIII, BDI, and CII,
allow random matrix ensembles 
with exact zero eigenvalue(s),
or ``zero-modes''.
(See Table \ref{tab: ivanov}.)
This in turn suggests the existence of
one-dimensional topological insulators in these 
classes\cite{FootnoteChiralOneDimension}.
To summarize, we have listed all 
 topological insulators as a function of symmetry and spatial 
 dimension in Table  \ref{tab: rmt}. The labels
 $\mathbb{Z}$ and $\mathbb{Z}_2$ used in this table 
 indicate whether the different topological sectors
 can be labeled by an integer or a $\mathbb{Z}_2$ 
 quantity, respectively.

\bigskip 

\subsection{Experimental implications}

The topological insulators discussed in this paper
can be realized in nature:
as explained in Sec.\ \ref{sec: 3D superconductors as a topological phase},
the fermionic sector of
the quasiparticles in
the
B-phase of liquid  $^{3}$He is an example of
the topological insulator in
the superconductor class DIII.
Also, 
an unconventional 
superconductor in a heavy-fermion compound, say,
could be a possible 
realization of a 3D topological phase,
possessing, e.g., a non-trivial
ground state degeneracy.
Another arena where these
novel 3D quantum states with exotic topological characteristics
may be realized experimentally
is that of cold atom systems
(tunable via the $p$-wave Feshbach resonance).
[The realization of a 2D topological phase
(the Moore-Read Pfaffian state or chiral 2D $p$-wave superconductor)
in cold atom systems has recently been discussed
in Refs.\ \onlinecite{Gurarie05} and \onlinecite{Gurarie06}.]
Unlike the B phase of liquid $^{3}$He, 
such a realization might allow us to go back and forth between
the weak and strong pairing phases of the topological insulators
by detuning.
Last but not the least,
strong correlations among electrons (or spins)
might \textit{spontaneously} give rise to these topological phases,
\cite{Raghu07}
by forming a non-trivial band
structure for fermionic excitations (e.g., spinons),
which can be explored by, e.g.,
slave-particle mean-field theories of a spin liquid.
\cite{comment Kitaev model}

One of the direct signatures of the non-trivial
topological characteristics of the 
quantum state in the three-dimensional
bulk is the appearance of gapless relativistic  
fermion modes at surfaces terminating
the bulk, 
which are stable to interactions and 
to disorder.
It should be possible to detect these surface states
using various 
experimental probes, 
such as tunneling/STM
probes, and
especially angle resolved photoemission spectroscopy
as already done successfully in
the Bismuth-Antimony alloys.
\cite{Hasan07}

Of particular interest are transport measurements,
as the 3D topological insulators
always possess
delocalized gapless modes
 propagating at 
their surface,
even in the presence of disorder.
For the 3D $\mathbb{Z}_2$ topological insulators
such surface modes
(in the symplectic symmetry class)
are predicted to be a perfect metal.
\cite{Ryu07,Nomura07}
Hence electrical transport measurements can be used to
determine if a specific insulating material is
a $\mathbb{Z}_2$ topological insulator or not,
which is a test independent of, say,
photoemission experiments, in which 
one counts the number of surface Dirac fermion
flavors.

We suggest that, for the superconducting classes DIII, AIII, and CI, 
it would be interesting to measure
either spin transport  
(for spin conserving symmetry classes, AIII and CI)
or thermal transport (for all the three classes)  properties
of the gapless delocalized surface modes.
The spin conductivity ($\sigma^{\mathrm{spin}}_{xx}$)
or the thermal conductivity divided by temperature ($\kappa^{\ }_{xx}/T$)
is, in the absence of disorder,
of order unity (in natural units)
because of
the vanishing density of states at zero energy:
when the Dirac cone is isotropic (i.e., the fermi velocity is the same
in all directions), 
the spin conductivity in classes AIII and CI is given \cite{Ludwig94}
by
$
\sigma^{\mathrm{spin}}_{xx}=
1/\pi
\times
s^2/h
$
per Dirac fermion, 
where $s=1/2$ is the spin ``charge'' 
carried by quasiparticles.
The thermal conductivity is then
given by
$
\kappa^{\ }_{xx}/T = 4\pi^2/(3
\sigma^{\mathrm{spin}}_{xx})
$
valid for all three classes AIII, DIII, and CI
(where for
classes AIII, and CI
this represents the
Wiedemann-Franz law).
For each 
surface Majorana fermion in class DIII, the thermal conductivity is
half the value obtained  
for a single Dirac fermion,
and the contributions from several flavors are additive.

The values for these
transport coefficients are completely 
robust against disorder.
This can be directly observed for the case of minimal number of surface
Dirac fermions;
Eq.\ (\ref{eq: surface Dirac AII}) with $V=0$ for class DIII,
Eq.\ (\ref{eq: single Dirac + gauge potential}) for class AIII,
and
Eq.\ (\ref{eq: surface Dirac CI}) for class CI.
For class DIII, 
there is simply no disorder potential allowed by symmetries.
While, as discussed, gauge type randomness is possible in class AIII
[random U(1) gauge field]
and
class CI 
[random SU(2) gauge field], 
it is known that they do not affect the transport coefficient.
\cite{Ludwig94,Tsvelik95,Ostrovsky06, susyWZW}
Thus, at the surface of
the 3D topological insulators in
 class DIII, AIII, and CI,
the transport coefficients (spin and thermal conductivities) are
temperature independent and universal.

\acknowledgments

The authors acknowledge helpful interactions with
Alexei Kitaev, Sung-Sik Lee,  Ashvin Vishwanath, 
and Grigory Volovik.
This work has been supported by the National Science Foundation (NSF) under
Grant No.\ PHY05-51164 and
by Grant-in-Aid for Scientific Research
from MEXT (Grant No.\ 16GS0219); it was also supported,
in part, by the NSF under DMR-0706140 (AWWL).
A.P.S.\ thanks the Swiss NSF for its financial support.

%\tableofcontents


\begin{thebibliography}{99}


\bibitem{Wen90}
X.-G.\ Wen,
Int.\ J. Mod.\ Phys.\ B \textbf{4}, 239 (1990).

\bibitem{Wen92}
X.-G.\ Wen,
Int.\ J. Mod.\ Phys.\ B \textbf{6}, 1711 (1992).

\bibitem{Wenbook}
X.-G.\ Wen,
\textit{Quantum Field Theory of Many-Body Systems},
Oxford University Press (2004).

\bibitem{Klitzing}
K.\ von Klitzing, G.\ Dorda, and M.\ Pepper,
Phys.\ Rev.\ Lett.\ \textbf{45}, 494 (1980).


\bibitem{Thouless82}
D.\ J.\ Thouless,
M.\ Kohmoto,
P.\ Nightingale,
and M.\ den Nijs,
Phys.\ Rev.\ Lett.\ \textbf{49}, 405 (1982).

\bibitem{Kohmoto}
M.\ Kohmoto,
Ann.\ Phys.\ (N.Y.) \textbf{160} 343 (1985).

\bibitem{KaneMele}
C.\ L.\ Kane and E.\ J.\ Mele,
Phys.\ Rev.\ Lett.\ \textbf{95}, 146802 (2005);
\textbf{95}, 226801 (2005).


\bibitem{Roy06}
%Title: On the $Z_2$ classification of Quantum Spin Hall Models
R.\ Roy,
\texttt{arXiv:cond-mat/0604211}.

\bibitem{Moore06}
%Title: Topological invariants of time-reversal-invariant band structures
%\texttt{cond-mat/0607314}.
J.\ E.\ Moore and L.\ Balents,
Phys.\ Rev.\ B \textbf{75}, 121306(R) (2007).

\bibitem{Roy3d}
%Three dimensional topological invariants for time reversal invariant
%Hamiltonians and the three dimensional quantum spin Hall effect
Rahul Roy,
\texttt{arXiv:cond-mat/0607531}.


\bibitem{Fu06_3Da}
L.\ Fu, C.\ L.\ Kane, and E.\ J.\ Mele,
Phys.\ Rev.\ Lett.\ \textbf{98}, 106803 (2007).
%\texttt{cond-mat/0607699}.

\bibitem{Fu06_3Db}
L.\ Fu and C.\ L.\ Kane,
Phys.\ Rev.\ B \textbf{76}, 045302 (2007).
%\texttt{cond-mat/0611341}.

\bibitem{Fu06}
%Title: Time Reversal Polarization and a Z_2 Adiabatic Spin Pump
L.\ Fu and C.\ L.\ Kane,
Phys.\ Rev.\ B \textbf{74}, 195312 (2006).
%\texttt{cond-mat/0606336}.

\bibitem{Bernevig05}
%Title: Quantum Spin Hall Effect
%cond-mat/0504147 [abs, ps, pdf, other] :
B.\ Andrei Bernevig and Shou-Cheng Zhang,
Phys.\ Rev.\ Lett.\ \textbf{96}, 106802 (2006).

\bibitem{Murakami06}
%arXiv:cond-mat/0607001 [ps, pdf, other]
%Title: Quantum Spin Hall Effect and Enhanced Magnetic Response by Spin-Orbit Coupling
Shuichi Murakami,
Phys.\ Rev.\ Lett.\ \textbf{97}, 236805 (2006).

\bibitem{Xiao-LiangQi08}
Xiao-Liang Qi, Taylor Hughes, and Shou-Cheng Zhang,
\texttt{arXiv:0802.3537}.


\bibitem{CongjunWu06}
%arXiv:cond-mat/0508273 [ps, pdf, other] :
%Title: The Helical Liquid and the Edge of Quantum Spin Hall Systems
Congjun Wu, B.\ Andrei Bernevig, and Shou-Cheng Zhang,
Phys.\ Rev.\ Lett.\ \textbf{96}, 106401 (2006).

\bibitem{Xu06}
C.\ Xu and J.\ E.\ Moore,
Phys.\ Rev.\ B \textbf{73}, 045322 (2006).

\bibitem{Takane04}
Y.\ Takane,
J.\ Phys.\ Soc.\ Jpn.\ \textbf{73}, 9 (2004);
                                 1430 (2004);
                                 2366 (2004).

\bibitem{Zirnbauer92}
Delocalization in the quasi-one-dimensional
symplectic class was discovered by
M.\ R.\ Zirnbauer,
Phys.\ Rev.\ Lett.\ \textbf{69}, 1584 (1992);
A.\ D.\ Mirlin, A.\ M\"uller-Groeling, and M.\ R.\ Zirnbauer,
Ann.\ Phys.\ (N.Y.) \textbf{236}, 325 (1994),
although the distinction between even and odd
number of channels was not appreciated then.

\bibitem{Ando98}
T. Ando and T. Nakanishi,  
J. Phys. Soc. Jpn.  \textbf{67} 1704 (1998);
T.\ Ando, T.\ Nakanishi, and R.\ Saito,
J. Phys. Soc. Jpn.  \textbf{67} 2857 (1998).

\bibitem{obuse2007}
H.~Obuse, A.~Furusaki, S.~Ryu, and C.~Mudry, 
Phys.\ Rev.\ B \textbf{76} 075301 (2007).

\bibitem{essin2007}
A.~M.~Essin and J.~E.~Moore, Phys.~Rev.~B \textbf{76}, 165307 (2007).

\bibitem{Ostrovsky07}
%cond-mat/0702115 [ps, pdf, other]
%Title: Quantum criticality and minimal conductivity in graphene with long-range disorder
P.\ M.\ Ostrovsky, I.\ V.\ Gornyi, and A.\ D.\ Mirlin,
Phys.\ Rev.\ Lett.\ \textbf{98}, 256801 (2007).


\bibitem{Ryu07}
%cond-mat/0702529 [ps, pdf, other]
%Title: Z2 topological term, the global anomaly, and the two-dimensional symplectic symmetry class of Anderson localization
Shinsei Ryu, Christopher Mudry, Hideaki Obuse, and Akira Furusaki,
Phys.\ Rev.\ Lett.\ \textbf{99}, 116601 (2007).

\bibitem{Bardarson07}
%arXiv:0705.0886 [ps, pdf, other]
%Title: Demonstration of one-parameter scaling at the Dirac point in graphene
J.\ H.\ Bardarson,
J.\ Tworzyd{\l}o,
P.\ W.\ Brouwer, 
and C.\ W.\ J.\ Beenakker,
Phys.\ Rev.\ Lett.\ \textbf{99}, 106801 (2007).


\bibitem{Nomura07}
%arXiv:0705.1607 [ps, pdf, other]
%Title: Topological delocalization of two-dimensional massless Dirac fermions
Kentaro Nomura, Mikito Koshino, and Shinsei Ryu,
Phys.\ Rev.\ Lett.\ \textbf{99}, 146806 (2007).

\bibitem{Haldane04}
%12. arXiv:cond-mat/0408417 [ps, pdf, other]
%Title: Berry Curvature on the Fermi Surface: Anomalous Hall Effect as a Topological Fermi-Liquid Property
F.\ D.\ M.\ Haldane,
Phys.\ Rev.\ Lett.\ \textbf{93}, 206602 (2004).


\bibitem{Xiao-LiangQi05}
%arXiv:cond-mat/0505308 [ps, pdf, other]
%Title: Topological quantization of the spin Hall effect in two-dimensional paramagnetic semiconductors
Xiao-Liang Qi, Yong-Shi Wu, and Shou-Cheng Zhang,
Phys.\ Rev.\ B \textbf{74}, 085308 (2006).


\bibitem{Bernevig-Taylor-Zhang06}
%arXiv:cond-mat/0611399 [ps, pdf, other]
%Title: Quantum Spin Hall Effect and Topological Phase Transition in HgTe Quantum Wells
B.\ Andrei Bernevig, Taylor L.\ Hughes, and Shou-Cheng Zhang,
Science \textbf{314}, 1757 (2006).

\bibitem{Koenig07}
%Title: Quantum Spin Hall Insulator State in HgTe Quantum Wells
Markus K\"onig,
Steffen Wiedmann, 
Christoph Bruene, 
Andreas Roth, 
Hartmut Buhmann, 
Laurens W. Molenkamp, 
Xiao-Liang Qi, 
and Shou-Cheng Zhang,
Science \textbf{318}, 766 (2007).
%Science, Sciencexpress 20 September 2007,
%\texttt{arXiv:0710.0582}.


\bibitem{Koenig08}
Markus K\"onig,
Hartmut Buhmann, 
Laurens W. Molenkamp, 
Taylor L. Hughes, 
Chao-Xing Liu,
Xiao-Liang Qi, and 
Shou-Cheng Zhang,
J.\ Phys.\ Soc.\ Jpn.\ \textbf{77}, 031007 (2008).
%\texttt{arXiv:0801.0901}.


\bibitem{Dai07}
%Title: Helical edge and surface states in HgTe quantum wells and bulk insulators
Xi Dai, Taylor L.\ Hughes, Xiao-Liang Qi, Zhong Fang, 
and Shou-Cheng Zhang, Phys.\ Rev.\ B \textbf{77}, 125319 (2008).
%\texttt{arXiv:0705.1516}.

\bibitem{Fukui06}
%Quantum spin Hall effect in three dimensional materials: Lattice computation of Z$_2$ topological invariants and its application to Bi and Sb
T.\ Fukui and Y.\ Hatsugai,
J.\ Phys.\ Soc.\ Jpn.\ \textbf{76}, 053702 (2007).
%\texttt{arXiv:cond-mat/0611423}.

\bibitem{Hasan07}
D.\ Hsieh, 
D.\ Qian, 
L.\ Wray, 
Y.\ Xia, 
Y.\ Hor, 
R.\ J.\ Cava, 
and M.\ Z.\ Hasan, 
Nature \textbf{452}, 970 (2008).
%970-974 (24 April 2008)  
%doi:10.1038/nature06843; 
%Nature (in press) (2008).



\bibitem{Zirnbauer96}
M.\ R.\ Zirnbauer,
J.\ Math.\ Phys.\ \textbf{37}, 4986 (1996).

\bibitem{Altland97}
A.\ Altland and M.\ R.\ Zirnbauer,
Phys.\ Rev.\ B \textbf{55}, 1142 (1997);
P. Heinzner, A. Huck Leberry, and M. R. Zirnbauer,
Commun.\ Math.\ Phys. \textbf{257}, 725 (2005).

\bibitem{footnote_PH}
In second quantized language the particle-hole transformation is
generally implemented on a non-interacting fermionic Hamiltonian
${\hat H} = 
\sum_{i,j} {\hat f}^\dagger_i  {\cal H }^{\ }_{i,j}  {\hat f}^{\ }_j$
by
${\hat f}^{\ }_i \leftrightarrow \sum_j U^*_{i j} {\hat f}^\dagger_j$.
($\hat{f}^{\dag}_i$ and
$\hat{f}^{\ }_i$
are canonical fermion operators
satisfying $\{ {\hat f}^\dagger_i, {\hat f}^{\ }_j\}=\delta_{i,j}$ ,
%$\{ {\hat f}^\dagger_i, {\hat f}^{\ }_j\}=\delta_{i,j}$ are canonical
%fermion operators,
${\cal H }_{i,j}$ is the single-particle Hamiltonian matrix,
and $U_{i j}$ denotes some unitary matrix).
Thus, under
the particle-hole transformation
$
{\hat H} 
\to
\sum_{i, i', j, j'} 
{\hat f}^{\ }_{i'}
{U^*}^\dagger_{i', i} 
{\cal H}_{i, j}
U^*_{j, j'}
{\hat f}^\dagger_{j'}
=
-
\sum_{j', i'}{\hat f}^\dagger_{j'}
%{\left(
(
{U^*}^\dagger {\cal H} U^*
)^*_{j', i'}
%\right )}^*_{j', i'}
{\hat f}_{i'}
+ \mathrm{tr}\, {\cal H}
$.
Therefore, the system is particle-hole invariant
if and only if there exists a unitary matrix $U$
so that
$ 
- U^\dagger 
( {\cal H} - {1\over 2} \mathrm{tr}\, {\cal H})^* U =
{\cal H} - {1\over 2} \mathrm{tr}\, {\cal H}$,
where the Hermiticity of ${\cal H}$ was used.
(We may ignore the constant shift $- {1\over 2} \mathrm{tr}\, {\cal H}$
of the energy eigenvalue.)
This condition then equals that in 
Eq. \ (\ref{eq: def epsilon and eta})
for $\epsilon_c=-1$
in Sec.\ \ref{sec: symmetry classification of non-interacting Hamiltonians}.
It can also be written in terms of the
complex conjugation operator $K$ which acts as 
$K^{-1} {\cal H} K = {\cal H}^*$.

\bibitem{footnote1}
Including parity symmetry (space inversion)
might also be interesting.
\cite{Fu06_3Db}

\bibitem{MehtaRandomMatrices}
e.g.: M. L. Mehta, \textit{Random Matrices};  third edition
(Elsevier, Amsterdam, 2004).

\bibitem{Furusaki99}
See, for example,
A. Furusaki, Phys.\ Rev.\ Lett.\ \textbf{82}, 604 (1999),
and references therein.

\bibitem{footnoteInsulator}
We shall denote the non-trivial topological phases 
in all five symmetry classes  by the term ``topological insulator'',
although in the context of 
superconducting systems (or superfluids) this
might be considered
a misnomer. In the case of the
BdG symmetry classes DIII and CI, 
as well as for
class AIII, when interpreted as a superconductor,\cite{Foster07}
the term ``insulator'' refers to the fact
that the BCS quasiparticles are fully gapped in the bulk
by the mean field pairing gap.
I.e., it refers to an insulating behavior
as manifested, e.g., by
thermal transport properties, or by spin transport
in superconductors
when spin is a good quantum number.
Alternatively, 
a superconductor or a superfluid
with non-trivial topological character
in BdG fermionic excitations
can be called topological superconductor
or topological superfluid, 
respectively. 


\bibitem{TRS in class AIII}
Although the TRS in class AIII
is only implicit 
when realized as a
random hopping problem of electrons with  sublattice symmetry,
class AIII 
can also be interpreted as an ensemble of
TR invariant triplet BdG Hamiltonians
of a superconductor
which is invariant under  a U(1) subgroup of the
SU(2) spin rotation symmetry,
such as rotation around the $z$-axis in spin space.
\cite{Foster07, FendleyKonik00}
See 
Sec.\
\ref{sec: symmetry classification of non-interacting Hamiltonians,BdG}.


\bibitem{Foster07}
%Metal-insulator transition from combined disorder and interaction effects in Hubbard-like electronic lattice models with random hopping
Matthew S.\ Foster and Andreas W.\ W.\ Ludwig,
\texttt{arXiv:}
\texttt{0710.0400v1}.

\bibitem{FendleyKonik00}
%On phase transitions in two-dimensional disordered systems
%arXiv:cond-mat/0003436v2 [cond-mat.supr-con]
P.\ Fendley and R.\ M.\ Konik,
Phys.\ Rev.\ B \textbf{62}, 9359 (2000).

\bibitem{balian1963} 
R.\ Balian and N.\ R.\ Werthamer, 
Phys.\ Rev.\ \textbf{131}, 1553 (1963).


\bibitem{VolovikBooks}
G.\ E.\ Volovik,
\textit{The Universe in a Helium Droplet}
(The International Series of Monographs on Physics, 117),
Oxford University. Press (2003);
G.\ E.\ Volovik,
\textit{Exotic Properties of Superfluid 3He}
(Series I N Modern Condensed Matter Physics, Vol 1),
World Scientific (1992).

\bibitem{Halperin82}
B.\ Halperin,
Phys.\ Rev.\ B \textbf{25}, 2185 (1982).



\bibitem{Callan85}
C.\ G.\ Callan Jr. and J.\ A.\ Harvey,
Nucl.\ Phys.\ B \textbf{250}, 427 (1985).


\bibitem{Fradkin86}
E.\ Fradkin, E.\ Dagotto, and D.\ Boyanovsky,
Phys.\ Rev.\ Lett. \textbf{57}, 2967 (1986).

\bibitem{Boyanovsky87}
D.\ Boyanovsky, E.\ Dagotto, and E.\ Fradkin,
Nucl.\ Phys.\ B \textbf{285}, 340 (1987).

\bibitem{Kaplan92}
%arXiv:hep-lat/9206013 [ps, pdf, other]
%Title: A Method for Simulating Chiral Fermions on the Lattice
D.\ B.\ Kaplan,
Phys.\ Lett.\ B \textbf{288}, 342 (1992).


\bibitem{SalomaaVolovik1988}
The fermion zero modes at an interface 
of two B phases of liquid $^{3}$He
with different momentum space topology
were discussed in
M.\ M.\ Salomaa and G.\ E.\ Volovik,
%Cosmiclike domain walls in superfluid 3B: Instantons and diabolical
%points in (k,r) space,
Phys.\ Rev.\ B \textbf{37}, 9298 (1988),
in terms of diabolical points in mixed
real and momentum space [$(k,r)$ space].

\bibitem{GrinevichandVolovik1988}
%Topology of gap nodes in superfluid $^3$He: $\pi_4$ homotopy group
%for $^3$He-B disclination,
P.\ G.\ Grinevich and G.\ E.\ Volovik,
J.\ Low Temp.\ Phys.\ \textbf{72}, 371 (1988).


\bibitem{WittenOldChernSimonsCFT}
%Quantum Field Theory and the Jones Polynomial
E.\ Witten,
Comm.\ Math.\ Phys.\ \textbf{121}, 351 (1989).

\bibitem{MooreRead91}
G.\ Moore and N.\ Read,
Nucl.\ Phys.\ B \textbf{360}, 362 (1991).

\bibitem{Hatsugai93}
Y.\ Hatsugai,
Phys.\ Rev.\ Lett.\ \textbf{71}, 3697 (1993).

\bibitem{Witten2007TwoPlusOneGravityAndAdS}
%Title: Three-Dimensional Gravity Revisited
E.\ Witten,
\texttt{arXiv:0706.3359}.


\bibitem{Bernard01}
%15. arXiv:cond-mat/0109552 [ps, pdf, other]
%Title: A Classification of random Dirac fermions
Denis Bernard and Andr\'e LeClair,
J.\ Phys.\ A \textbf{35}, 2555 (2002).

\bibitem{NoKnowledgeofBernardLeClair}
Though we stress no previous knowledge of
%this work
Ref. [\onlinecite{Bernard01}]
 is required of the reader to understand
our paper.



\bibitem{Read00}
N.\ Read and D.\ Green,
Phys.\ Rev.\ B \textbf{61}, 10267 (2000).



\bibitem{Diamantini1996}
%arXiv:hep-th/9511168 [ps, pdf, other]
%Title: Gauge Theories of Josephson Junction Arrays
M.\ C.\ Diamantini, P.\ Sodano,  and C.\ A.\ Trugenberger,
Nucl.\ Phys.\ B \textbf{474},  641 (1996).

\bibitem{Hansson04}
T.\ H.\ Hansson,
V.\ Oganesyan,
and
S.\ L.\ Sondhi,
Ann.\ Phys.\ (N.Y.) \textbf{313}, 497 (2004).
%\texttt{arXiv:cond-mat/0404327}.

\bibitem{Diamantini2006}
%arXiv:cond-mat/0410208 [ps, pdf, other]
%    Title: Superconducting Topological Fluids in Josephson Junction Arrays
M.\ Cristina Diamantini, Pasquale Sodano,  and Carlo A.\ Trugenberger,
J.\ Phys.\ A: Math.\ Gen.\ \textbf{39}, L253 (2006).


\bibitem{Oshikawa06}
%Topological degeneracy of non-Abelian states for dummies
M.\ Oshikawa, Yong Baek Kim, K.\ Shtengel,
C.\ Nayak, and S.\ Tewari,
Ann.\ Phys.\ (N.Y.) \textbf{322}, 1477 (2007).
%\texttt{arXiv:cond-mat/0607743}.


\bibitem{Halperin83}
B.\ I.\ Halperin, 
Helv.\ Phys.\ Acta \textbf{56}, 75 (1983).

\bibitem{HaldaneRezayi88}
F.\ D.\ M.\ Haldane
and 
E.\ H.\ Rezayi, 
Phys.\ Rev.\ Lett.\ \textbf{60}, 956 (1988);
\textbf{60}, 1886 (1988).

%%%%%%%%%%%%%%%%%%%%%%%%%%%%%%%%%%%%%%%%%%%%%%%%%%%%%%


\bibitem{Bauer04}
E.\ Bauer,
G.\ Hilscher,
H.\ Michor,
Ch.\ Paul,
E.\ W.\ Scheidt,
A.\ Gribanov,
Yu.\ Seropegin,
H.\ No\"el,
M.\ Sigrist,
and P.\ Rogl,
Phys.\ Rev.\ Lett.\ \textbf{92}, 027003 (2004).


\bibitem{SenguptaRoySC2}
%arXiv:cond-mat/0604217 [ps, pdf, other]
%Title: Spin-Hall effect in triplet chiral superconductors and graphene
K.\ Sengupta, Rahul Roy, and Moitri Maiti,
Phys.\ Rev.\ B \textbf{74}, 094505 (2006).

\bibitem{RoySC}
%Title: Topological invariants of time reversal invariant superconductors
Rahul Roy,
\texttt{arXiv:cond-mat/0608064}.

\bibitem{RoySF2008}
%Title: Topological superfluids with time reversal symmetry
Rahul Roy,
\texttt{arXiv:0803.2868}.

\bibitem{Xiao-LiangQiTopSC08}
%Topological Superconductivity and Superfluidity
Xiao-Liang Qi, 
Taylor L.\ Hughes, 
Srinivas Raghu, and
Shou-Cheng Zhang, 
\texttt{arXiv.0803.3614}.

\bibitem{SenthlFisherBalentsNayak1998}
%arXiv:cond-mat/9808001 [ps, pdf, other]
%Title: Quasiparticle transport and localization in high-T_c superconductors
T.\ Senthil, Matthew P.\ A.\ Fisher, Leon Balents, and Chetan Nayak,
Phys. \ Rev. \ Lett.\ \textbf{81}, 4704 (1998).


\bibitem{SenthilFisher2000}
%Quasiparticle localization in superconductors with spin-orbit scattering
T.\ Senthil and Matthew P. \ A.\ Fisher,
Phys.\ Rev.\ B \textbf{61}, 9690 (2000);
N. Read and Andreas W. W. Ludwig, Phys. Rev. B  \textbf{63}, 024404 (2000);
J. T. Chalker, N. Read, V. Kagalovsky, B. Horovitz, Y. Avishai, and A. W. W. Ludwig,
Phys. Rev. B \textbf{65}, 012506 (2001).





\bibitem{SenthilMarstonFisher}
T. Senthil, J. B. Marston, and M. P. A. Fisher,
Phys.\ Rev.\ B \textbf{60}, 4245 (1999);
Ilya A. Gruzberg, Andreas W. W. Ludwig, and N. Read, Phys. Rev. Lett.
\textbf{82}, 4524 (1999).




\bibitem{Altland02}
See for example,
%arXiv:cond-mat/0006362,
%\textit{Theories
%of Low-Energy Quasi-Particle States in Disordered d-Wave Superconductors},
A.\ Altland, B.\ D.\ Simons, and M.\ R.\ Zirnbauer,
Phys.\ Rep.\ \textbf{359}, 283 (2002).


%%%%%%%%%%%%%%%%%%%%%%%%%%%%%%%%%%%%%%%%%%%%%%%%%%%%%%%%%%%%%%%%%%%%%%%%%%%%

\bibitem{Avron83}
%Homotopy and Quantization in Condensed Matter Physics
J.\ E.\ Avron,  R.\ Seiler, and  B.\ Simon,
Phys.\ Rev.\ Lett.\ \textbf{51}, 51 (1983).

\bibitem{Novikov}
See, for example, D.\ B.\ Fuchs and O.\ Ya.\ Viro, \emph{Topology II}, edited by V.\ A.\ Rokhlin 
and S.\ P.\ Novikov, Encyclopaedia of Mathematical Sciences Vol.\ 24 (Springer, New York, 2004).
%See for example,
%S.\ P.\ Novikov and V.\ A.\ Rokhlin (Eds.),
%\textit{Topology II}, Encyclopaedia of Mathematical Sciences vol.\ 24,
%Springer (2004).

\bibitem{Nakahara03}
M.\ Nakahara,
\textit{Geometry, Topology, and Physics},
(second edition, Institute of Physics Publishing, Bristol, 2003).

\bibitem{Comment 3DIQHE}
This is not to say that the 3D IQHE on \textit{a lattice}, discussed by
Kohmoto, Halperin, and Wu, 
Phys. Rev. B \textbf{45},  13488 (1992), 
is not possible: in the above work by Kohmoto \textit{et al.},
quantum ground states constructed from filled 3D Bloch states
are characterized by a triplet of Chern numbers, each describing
the winding of a map from the 2D torus, which is a subspace of
3D BZ, onto $G_{m,m+n} ( \mathbb{C} )$. Hence, the 3D IQHE is
essentially a layered version of the 2D IQHE. Similarly, in symmetry
class AII, there exists a 3D topological state, which consists of  layered
2D $\mathbb{Z}_2$ topological quantum states, and which has been
termed ``weak topological insulator'' (see Ref.\ \onlinecite{Fu06_3Da}).
By analogy, we argue that also in symmetry classes D, C, and DIII  (see Table I)
there exists a layered version of the 2D topological quantum states.


%
%\bibitem{Kohmoto-Halperin-Wu92}
%%Diophantine equation for the three-dimensional quantum Hall effect
%M.\ Kohmoto, B.\ I.\ Halperin, and Y.-S. Wu,
%Phys.\ Rev.\ B \textbf{45}, 13488 (1992).

\bibitem{Moore2008}
J.\ E.\ Moore, Y.\ Ran, and X.\ -G.\ Wen, 
\texttt{arXiv:0804.4527}.

\bibitem{SSLee07}
%arXiv:0708.1639 [ps, pdf, other]
%A many-body generalization of the Z2 topological invariant for the quantum spin Hall effect
Sung-Sik Lee and Shinsei Ryu,
Phys.\ Rev.\ Lett.\ \textbf{100}, 186807 (2008).
%\texttt{arXiv:0708.1639}.

\bibitem{comment Roy and Sengupta}
Topological insulators in TR invariant BdG classes
have been discussed 
in terms of the $\mathbb{Z}_2$ number in Refs.\ 
\onlinecite{SenguptaRoySC2},
\onlinecite{RoySC}, 
\onlinecite{RoySF2008}, and 
\onlinecite{Xiao-LiangQiTopSC08},
%where a classification in terms of
%the $\mathbb{Z}_2$ number was discussed
thereby emphasizing the fermion number parity
in the ground state
(See also Sec.\ \ref{sec: 3D superconductors as a topological phase}).
Here, however, 
we find that an additional discrete symmetry (i.e., PHS),
allows to define an integral winding number $\nu$, which protects
an arbitrary number (an arbitrary even number for class CI) 
of surface Dirac (Majorana for class DIII) fermion states
against the opening of a gap.

 
\bibitem{Ryu01}
%arXiv:cond-mat/0112197 [ps, pdf, other]
%Title: Topological Origin of Zero-Energy Edge States in Particle-Hole Symmetric Systems
S.\ Ryu and Y.\ Hatsugai,
Phys.\ Rev.\ Lett.\ \textbf{89}, 077002 (2002).

\bibitem{Kitaev00}
%Title: Unpaired Majorana fermions in quantum wires
Alexei Kitaev,
in the proceedings of the {\it
Mesoscopic
And Strongly Correlated Electron Systems} conference
(9-16 July 2000, Chernogolovka, Moscow Region, Russia),
\texttt{arXiv:cond-mat/0010440}.

\bibitem{Kitaev05}
%arXiv:cond-mat/0506438 [ps, pdf, other]
%Title: Anyons in an exactly solved model and beyond
Alexei Kitaev,
Ann.\ Phys.\ (N.Y.) \textbf{321}, 2 (2006).

\bibitem{RyuEE06}
%arXiv:cond-mat/0601237 [ps, pdf, other]
%Title: Entanglement entropy and the Berry phase in solid states
S.\ Ryu and Y.\ Hatsugai,
Phys.\ Rev.\ B \textbf{73}, 245115 (2006).






\bibitem{Haldane88}
F.\ D.\ M.\ Haldane,
Phys.\ Rev.\ Lett.\ \textbf{61}, 2015 (1988).

\bibitem{Niu85}
Q.\ Niu, D.\ J.\ Thouless, and Y.-S.\ Wu,
Phys.\ Rev.\ B \textbf{31}, 3372 (1985).


%%%%%%%%%%%%%%%%%%%%%%%%%%%%%%%%%%%%%%%%%%%%%%%%%%%

\bibitem{Fendley00}
%arXiv:cond-mat/0006360 [ps, pdf, other]
%Title: Critical points in two-dimensional replica sigma models
See, for example, 
Paul Fendley, 
in \emph{New Theoretical Approaches to Strongly Correlated Systems}, 
edited by A.\ M.\ Tsvelik (Kluwer Academic Publishers, The Netherlands, 2001); arXiv:cond-mat/0006360.
%e.g. Paul Fendley,
%Lecture at the NATO Advanced Study Institute/EC Summer School on New
%Theoretical Approaches to Strongly Correlated Systems, Newton
%Institute, Cambridge, UK, April 10-20, 2000;
%\texttt{arXiv:cond-mat/0006360}.



\bibitem{Fendley01}
P.\ Fendley,
Phys.\ Rev.\ B \textbf{63}, 104429 (2001).

%%%%%%%%%%%%%%%%%%%%%%%%%%%%%%%%%%%%%%%%%%%%%%%%%%%

\bibitem{commentBernevigAndChen}
In B.\ Andrei Bernevig and Han-Dong Chen,
\texttt{arXiv:}
\texttt{cond-mat/0611766v1},
a topological integer was introduced
in the context of the 
$\mathbb{Z}_2$ topological insulator in 3D,
for a \textit{subset} of $4\times 4$ Hamiltonians in
the symplectic symmetry class. 
The condition that selects this subset
turns out to be PHS.
Hence, 
the relevant symmetry classes studied 
in the above work can be identified as
class DIII or class AIII,
rather than the symplectic symmetry class.


\bibitem{Deser82}
S.\ Deser, R.\ Jackiw, and S.\ Templeton,
Ann.\ Phys.\ (N.Y.) \textbf{140}, 372 (1982).

\bibitem{Ludwig94}
A.\ W.\ W.\ Ludwig,
M.\ P.\ A.\ Fisher,
R.\ Shankar,
and G.\ Grinstein,
Phys.\ Rev.\ B \textbf{50}, 7526 (1994).

\bibitem{Berry84}
M.\ V.\ Berry, 
Proc.\ R.\ Soc.\ London Ser.\ A \textbf{392}, 45 (1984).

\bibitem{Chruscinski}
D.\ Chru\'sci\'nski and A.\ Jami{\l}okoski,
\textit{Geometric Phases in Classical and Quantum Mechanics},
Progress in mathematical physics Volume 36,
Birkh\"auser Boston (2004).

\bibitem{Biswas89}
S.\ N.\ Biswas,
Phys.\ Lett.\ B \textbf{228}, 440 (1989).

\bibitem{Arodz89}
H.\ Arodz and A.\ Babiuch,
Acta Phys.\ Pol.\ B \textbf{20}, 579 (1989).

\bibitem{Hatsugai04}
Y.\ Hatsugai, S.\ Ryu, and M.\ Kohmoto,
Phys.\ Rev.\ B \textbf{70}, 054502 (2004).

\bibitem{Chyh-Hong04}
Chyh-Hong Chern,
Han-Dong Chen,
Congjun Wu,
Jiang-Ping Hu,
and
Shou-Cheng Zhang,
Phys.\ Rev.\ B \textbf{69}, 214512 (2004).

\bibitem{Wilczek and Zee}
F.\ Wilczek and A.\ Zee,
Phys.\ Rev.\ Lett.\ \textbf{52}, 2111 (1984).


\bibitem{Guruswamy00}
S.\ Guruswamy, A.\ LeClair, and A.\ W.\ W.\ Ludwig,
Nucl.\ Phys.\ B \textbf{583}, 475 (2000).


%%%%%%%%%%%%%%%%%%%%%%%%%%%%%%%%%%%%%%%%%%%%%%%%%%%%%

\bibitem{Golterman92}
%arXiv:hep-lat/9209003 [ps, pdf, other]
%Title: Chern-Simons Currents and Chiral Fermions on the Lattice
Maarten Golterman, Karl Jansen, and David Kaplan,
Phys.\ Lett.\ B \textbf{301}, 219 (1993).

%\bibitem{Ryu07LandauerBCFT}
%S.\ Ryu, C.\ Mudry, A.\ Furusaki, and A.\ W.\ W.\ Ludwig,
%%\textit{Landauer conductance and twisted boundary conditions for Dirac fermions in two space dimensions},
%Phys.\ Rev.\ B \textbf{75}, 205344 (2007).

\bibitem{Redlich84}
A.\ N.\ Redlich,
Phys.\ Rev.\ D \textbf{29}, 2366 (1984).

\bibitem{Dunne99}
%Aspects of Chern-Simons Theory
See for example,
Gerald V. Dunne,
Les Houches Lectures 1998,
\texttt{arXiv:hep-th/9902115}.


\bibitem{KnizhnikZamolodchikov}
V.\ G.\ Knizhnik and A.\ B.\ Zamolodchikov,
Nucl.\ Phys.\ B \textbf{247}, 83 (1984).


%%%%%%%%%%%%%%%%%%%%%%%%%%%%%%%%%%%%%%%%%%%%%%%%

\bibitem{Misirpashaev95}
T.\ Sh.\ Misirpashaev and G.\ E.\ Volovik,
Physica B \textbf{210}, 338 (1995).

\bibitem{Sato03}
M.\ Sato,
Phys.\ Lett.\ B \textbf{575}, 126 (2003).


%%%%%%%%%%%%%%%%%%%%%%%%%%%%%%%%%%%%%%%%%%%%%%%%%%%%%%%%%

\bibitem{LeeRamakrishnan} 
Patrick A.\ Lee
and
T.\ V.\ Ramakrishnan,
Rev.\ Mod.\ Phys.\ \textbf{57}, 287 (1985).

\bibitem{Efetov97} 
K.\ Efetov, 
\textit{Supersymmetry in disorder and chaos}, 
(Cambridge University Press, Cambridge, 1997).

\bibitem{commentPruisken}
While the topological term of Pruisken type is possible 
for classes A, C, and D in two dimensions, 
such a term need not necessarily 
be realized 
at a surface of a 3D topological insulator.
What distinguishes the Pruisken terms from
$\mathbb{Z}_2$ topological terms or the WZW terms
is their tunability:
the Pruisken term depends on one parameter (topological angle),
which can be tuned by changing microscopic details.


\bibitem{Brouwer05}
P.\ W.\ Brouwer, A.\ Furusaki, C.\ Mudry, and S.\ Ryu,
BUTSURI \textbf{60}, 935 (2005);
\texttt{arXiv:cond-mat/0511622}.

\bibitem{Caselle06}
%``Symmetric space description of carbon nanotubes'',
M.\ Caselle and U.\ Magnea,
%J.\ Stat.\ Mech.\ 0601 (2006) P013.
J. Stat.\ Mech.: Theory Exp.\ \textbf{2006}, P01013.

\bibitem{Gruzberg05}
%arXiv:cond-mat/0412413 [ps, pdf, other]
%``Localization in disordered superconducting wires with broken spin-rotation symmetry'',
Ilya A.\ Gruzberg, N.\ Read, and Smitha Vishveshwara,
Phys.\ Rev.\ B \textbf{71}, 245124 (2005).

\bibitem{Ivanov99}
%arXiv:cond-mat/9911147 [ps, pdf, other]
%Title: The energy-level statistics in the core of a vortex in a p-wave superconductor
D.\ A.\ Ivanov,
\texttt{arXiv:}\texttt{cond-mat/9911147};
%26. arXiv:cond-mat/0103137 [ps, pdf, other]
%Title: The supersymmetric technique for random-matrix ensembles with zero eigenvalues
D.\ A.\ Ivanov,
J.\ Math.\ Phys.\ \textbf{43}, 126 (2002).



\bibitem{FootnoteChiralOneDimension}
Note that the three classes
AIII, BDI, and CII possess sublattice symmetry (SLS),
and can thus be realized as nearest neighbor hopping models
on a 1D lattice (which is always bipartite). There are two dimerized states, and 
for a finite lattice
one of them has a zero-mode ``edge state''  at each of the two boundaries.
The topological integer specifies the occupation number of
such a zero-mode ``edge state''.



\bibitem{Gurarie05}
V.\ Gurarie, L.\ Radzihovsky,
and A.\ V.\ Andreev,
Phys.\ Rev.\ Lett.\ \textbf{94}, 230403 (2005).

\bibitem{Gurarie06}
%arXiv:cond-mat/0611022 [ps, pdf, other]
%Title: Resonantly-paired fermionic superfluids
V.\ Gurarie and L.\ Radzihovsky,
Ann.\ Phys.\ (N.Y.) \textbf{322}, 2 (2007).


\bibitem{Raghu07}
%Title: Topological Mott Insulators
S.\ Raghu, Xiao-Liang Qi, C.\ Honerkamp, 
and Shou-Cheng Zhang,
Phys.\ Rev.\ Lett.\ \textbf{100}, 156401 (2008).
%\texttt{arXiv:0710.0030}.

\bibitem{comment Kitaev model}
It is possible to design an interacting spin model
for which a fermionic many-body state constructed from
a slave-particle mean field Hamiltonian (projective construction) 
is an exact ground state.
\cite{Wenbook,Kitaev05} 
Indeed, in Ref.\ \onlinecite{Kitaev05}, 
Kitaev discussed a spin-1/2 model on the honeycomb lattice, 
whose ground state is constructed from non-interacting Majorana fermions
and
which lies
in the universality class of the Moore-Read Pfaffian state. 
Following the spirit of Ref.~\onlinecite{Kitaev05}
we can construct 
an exactly solvable spin-3/2 model of Kitaev type 
on the diamond lattice
whose ground state can be obtained from a fermionic ground state
of a class DIII topological insulator. 


\bibitem{Tsvelik95}
%cond-mat/9409039 [abs, ps, pdf, other] :
%Title: An Exactly Solvable Model of Fermions with Disorder
A.\ M.\ Tsvelik,
Phys.\ Rev.\ B \textbf{51}, 9449 (1995).

\bibitem{Ostrovsky06}
%Title: Electron transport in disordered graphene
P.\ M.\ Ostrovsky, I.\ V.\ Gornyi, and A.\ D.\ Mirlin,
Phys.\ Rev.\ B \textbf{74}, 235443 (2006).
%\texttt{cond-mat/0609617}.

\bibitem{susyWZW}
When using supersymmetry method for disorder averaging,
the resulting theories are, in general, WZW models
on GL($1|1$) and Osp($2|2$) supergroup manifolds, possessing Kac-Moody current
algebra symmetry.

 
\end{thebibliography}
\end{document}